\documentclass[pdflatex,sn-mathphys-num]{sn-jnl}% Math and Physical Sciences Numbered Reference Style

\usepackage{graphicx}%
\usepackage{multirow}%
\usepackage{amsmath,amssymb,amsfonts}%
\usepackage{amsthm}%
\usepackage{mathrsfs}%
\usepackage[title]{appendix}%
\usepackage{textcomp}%
\usepackage{manyfoot}%
\usepackage{booktabs}%
\usepackage{subcaption}%
\usepackage{bm}
\usepackage{ragged2e}

\usepackage{array}
\newcolumntype{+}{>{\global\let\currentrowstyle\relax}}
\newcolumntype{^}{>{\currentrowstyle}}

\renewenvironment{table}[1][]%
{\tableorg[#1]%
\tablebodyfont%
\renewcommand\footnotetext[2][]{{\removelastskip\vskip3pt%
\let\tablebodyfont\tablefootnotefont%
\hskip0pt\if!##1!\else{\smash{$^{##1}$}}\fi##2\par}}%
}{\endtableorg}

\usepackage{float}%
\usepackage{algorithmic}
\usepackage{algorithm}%
\usepackage{listings}%
\usepackage{longtable}
\usepackage[table]{xcolor}

\theoremstyle{thmstyleone}%

\theoremstyle{thmstyletwo}%

\theoremstyle{thmstylethree}%

\begin{document}

\title[Simulation-Based Sample Size Estimation for Clinical Prediction Models: The \texttt{pmsims} Package]%
{Adaptive Gaussian Process Search for Simulation-Based Sample Size Estimation in Clinical Prediction Models: Validation of the \texttt{pmsims} \textsf{R} Package}

\author*[1,2]{\fnm{Oyebayo Ridwan} \sur{Olaniran}}\email{ridwan.olaniran@kcl.ac.uk}

\author[1,2]{\fnm{Diana} \sur{Shamsutdinova}}

\author[1]{\fnm{Sarah} \sur{Markham}}

\author[1]{\fnm{Felix} \sur{Zimmer}}

\author[1,2]{\fnm{Daniel} \sur{Stahl}}

\author[1,2]{\fnm{Gordon} \sur{Forbes}}
\equalcont{These authors contributed equally to this work.}

\author[1]{\fnm{Ewan} \sur{Carr}}
\equalcont{These authors contributed equally to this work.}

\affil*[1]{\orgdiv{Department of Biostatistics and Health Informatics}, \orgname{King's College London}, \orgaddress{\city{London}, \country{United Kingdom}}}

\affil[2]{\orgdiv{NIHR Biomedical Research Centre}, \orgname{Maudsley NHS Trust}, \orgaddress{\city{London}, \country{United Kingdom}}}

\abstract{
\textbf{Background}

\noindent Determining an adequate sample size is essential for developing reliable and generalisable clinical prediction models, yet practical guidance on selecting appropriate sample size calculation methods remains limited. Existing analytical and simulation-based tools impose restrictive assumptions and focus on mean-based criteria. We present and validate \texttt{pmsims}, an \textsf{R} package that uses Gaussian process (GP) surrogate modelling to provide a flexible and efficient simulation-based framework for sample size determination across a broad range of prediction modelling contexts.\\

\textbf{Methods}

\noindent We conducted a comprehensive simulation study with two aims. Aim~1 compared three search engines implemented in \texttt{pmsims}, a GP surrogate-based adaptive procedure (\texttt{gp}), a deterministic bisection method (\texttt{bisection}), and a hybrid GP-bisection approach (\texttt{gp-bs}), across binary, continuous, and survival outcomes. Scenarios varied outcome prevalence or event rate, predictor dimensionality ($p = \{10, 100\}$), target performance metric (discrimination and calibration slope), aggregation criterion (mean vs 80\% assurance), and total simulation budget ($B = 200-2000$). Each scenario was replicated 100 times; estimator stability was assessed via the coefficient of variation (CV). Aim~2 benchmarked the best-performing \texttt{pmsims} engine against \texttt{pmsampsize} (analytical) and \texttt{samplesizedev} (simulation-based) across a wider range of realistic scenarios, evaluating recommended sample sizes, computational time, and achieved model performance on independent validation datasets of 30{,}000 observations.\\

\textbf{Results}

\noindent The GP-based search engine consistently yielded the most stable sample size estimates (lowest CV) across all outcome types, ranking highest in 9/12 outcome-aggregation metric configurations. Its advantage was most pronounced in low-signal, high-dimensional settings, and was accentuated with $\kappa = 20$ replications per evaluation and a budget of $B \approx 1000$, after which gains were minimal. In benchmark comparisons, \texttt{pmsims} (mean) achieved performance deviations within $\pm 1\%$ of the prespecified target across binary, continuous, and survival outcomes, comparable to \texttt{samplesizedev} and substantially outperforming \texttt{pmsampsize} in high-discrimination settings (deviations up to $-9.84\%$).\\

\textbf{Conclusions}

\noindent The \texttt{pmsims} package, using the GP-based search engine with $\kappa \ge 20$ replications per evaluation and a budget of $B \approx 1000$, provides a computationally efficient and flexible framework for principled sample size planning in clinical prediction modelling. It reliably achieves performance targets across a wide range of scenarios while requiring fewer model evaluations than non-adaptive simulation approaches, offering a compelling alternative to both analytical formulae and exhaustive simulation-based search.
}


\maketitle

\section{Introduction}

Determining an adequate sample size is an essential prerequisite for developing stable, reliable, and generalisable clinical prediction models. Development samples that are too small can produce overfitted models with poor calibration and reduced discrimination when applied to new data, undermining clinical utility and potentially causing harm in decision-making contexts~\cite{VanCalster2016}.

In prediction modelling, sample size estimation seeks the minimum development sample size needed for model performance to meet or exceed a prespecified threshold on a chosen metric~\cite{Diana2026pmsims}. Targets can be defined using either a `mean' criterion, where expected performance exceeds the threshold, or an `assurance' criterion, where performance exceeds the threshold with high probability (e.g., 80\%). The assurance approach explicitly accounts for training-sample variability and is therefore more robust to uncertainty in model development~\cite{sadatsafavi2025bayesian, pavlou2025sample}.

Analytic tools have been developed to estimate minimum sample sizes for traditional regression-based prediction models. For example, the \texttt{pmsampsize} R library and Stata modules target an acceptable degree of overfitting (via shrinkage) or ensure sufficiently precise estimation of the average outcome in the population~\cite{Riley2020}. However, these approaches are limited to linear or generalised linear models estimated using maximum likelihood and are not generalisable to a wider class of machine learning models or alternative estimation methods such as penalised maximum likelihood. Moreover, they rely on assumptions that are often unmet in practice. In particular, they typically presume a correctly specified linear predictor, normally distributed covariates, and properties of maximum likelihood estimation---assumptions that can be violated in real-world clinical data characterised by non-linearities, interactions, and complex correlation structures~\cite{VanderPloeg2014}. A further limitation of existing analytical methods is that they focus on the `mean' criterion, which does not capture variability in performance arising from finite training samples. Consequently, a sample size that meets a mean target may still imply a substantial probability of underperformance in the target population---risk not captured by closed-form formulae. Simulation studies have also shown analytic approaches may underestimate the required sample size when model strength is high (C-statistic $\ge$ 0.8) \cite{pavlou2024evaluation}.

To address the limitations of analytic approaches, simulation-based sample size tools have been developed. \texttt{samplesizedev} is a simulation-based R package that estimates sample sizes for binary and survival outcomes, quantifying variability in model performance metrics and supporting probability-based assessment of achieving target performance across training samples (the `assurance' criterion)~\cite{pavlou2025sample}. More recently, Bayesian approaches have also been proposed to support probability-based (assurance-type) sample size criteria for prediction modelling~\cite{sadatsafavi2025bayesian, riley2025general, pavlou2025sample}. \texttt{samplesizedev} only supports logistic and Cox regression models, simulates predictors independently from standard normal distributions, and is not readily extendable to other models or data generators. Other approaches to sample size determination include \textit{learning curve methods}, which extrapolate model performance from pilot data; and \textit{hybrid approaches}, in which estimated learning curves are summarised as approximate analytical formulae parametrised by the characteristics of the datasets used to derive them. These approaches are reviewed in~\cite{Diana2026pmsims}.

Despite these advances, there is a lack of practical, accessible tools for assurance-type sample size determination accommodating diverse model specifications and realistic data-generating mechanisms. This gap presents researchers with an unhelpful choice:\ rely on simplified approaches that may not reflect the complexities of the intended application or omit sample size justifications altogether. Consistent with this, systematic reviews show that most prediction modelling studies do not report an explicit sample size justification~\cite{Dhiman2023}.

In this paper, we present \texttt{pmsims}, an R package that provides a flexible simulation-based framework for estimating minimum sample sizes for developing clinical prediction models. The package addresses two core limitations of existing analytical approaches: limited flexibility in model and data-generating specifications, and the computational burden of na\"{\i}ve simulation-based searches over large sample size spaces~\cite{Diana2026pmsims}.

The \texttt{pmsims} package incorporates two key innovations. First, it uses Gaussian process (GP) surrogate modelling of the sample size--performance relationship \cite{zimmer2024sample,wilson2021efficient}. Rather than exhaustively evaluating every candidate sample size via computationally expensive model fitting, the GP-based engine uses a Gaussian process surrogate model to approximate performance as a function of sample size. It iteratively selects the next sample sizes to evaluate by choosing points where the current Gaussian process shows the highest posterior uncertainty (or where an acquisition function that balances predicted performance and uncertainty is maximised), runs a limited number of Monte Carlo simulations at those points, updates the GP posterior with the new results, and repeats until the surrogate provides a smooth, reliable approximation of the entire performance curve \cite{Diana2026pmsims}. This approach, grounded in Bayesian optimisation principles, dramatically reduces the number of required model fits compared to traditional grid search or bisection algorithms \cite{zimmer2024sample}. Second, \texttt{pmsims} provides a flexible framework for sample size calculations across any model type, data-generating mechanism, and performance metric of interest. This generality moves beyond the narrower focus of existing tools, enabling researchers to tailor calculations to their specific prediction modelling context rather than fitting their problem to available software.

This study presents the comprehensive validation of the \texttt{pmsims} package (version~0.5.0) through two complementary objectives. First, we evaluate the performance and computational efficiency of the GP-based engine against traditional simulation-based search algorithms. Second, we assess empirical agreement between sample size estimates from \texttt{pmsims} and those from established analytical and simulation-based tools, specifically \texttt{pmsampsize} (implementing the Riley et al.\ methodology for logistic, linear and survival outcomes)~\cite{pate2023minimum} and \texttt{samplesizedev} (a simulation-based approach motivated by limitations of analytical formulae in certain scenarios)~\cite{pavlou2024evaluation}, across a broad spectrum of realistic prediction modelling scenarios.

\section{The pmsims workflow}

The \texttt{pmsims} package implements a flexible simulation framework for sample size estimation, described by three components: (i) a data-generating process, (ii) a model-fitting procedure, and (iii) a performance metric. These components are combined with a search engine to estimate the minimum sample size $n$ that satisfies a prespecified performance criterion, either on average (the `mean' criterion) or with a specified level of certainty (the `assurance' criterion). Further details of the conceptual framework are provided in~\cite{Diana2026pmsims}.

Although \texttt{pmsims} supports arbitrary combinations of data-generating processes, model-fitting procedures, and performance criteria, we restricted the scope of this validation study to enable direct comparison with existing sample size tools. Specifically, we focused on binary, continuous, and survival outcomes; simple data-generating processes; and logistic regression, linear regression, and proportional hazards models. These are defined formally below and are provided in the package as pre-defined functions.

\subsection{Data-generating process}

Let \( \mathbf{X} = (X_1, X_2, \dots, X_p) \) be a \( p \)-dimensional vector of predictors, where \( p = p_{\text{signal}} + p_{\text{noise}} \). The data-generating process is defined by the joint distribution \( P(\mathbf{X}, Y) \), where the outcome \( Y \) is generated conditionally on \( \mathbf{X} \). For each outcome type, the linear predictor
\[
\eta = \mathbf{X}^T \boldsymbol{\beta} + \beta_0
\]
is constructed, where \( \boldsymbol{\beta} \) contains coefficients for the signal predictors (all set to a common value \( \beta_{\text{signal}} \)) and zeros for noise predictors. This structure reflects standard assumptions in simulation studies for evaluating prediction models \cite{van2019calibration}. The current version of the \texttt{pmsims} package (0.5.0) provides three data-generating processes:

\vspace{0.6\baselineskip}
\noindent\begin{minipage}{\linewidth}
\begin{tabular}{@{}p{0.15\linewidth} >{\RaggedRight\arraybackslash}p{0.83\linewidth} @{}}
\textbf{Binary outcomes} &
\( Y \sim \text{Bernoulli}(\pi) \), where \( \pi = \text{logit}^{-1}(\eta) \). The baseline probability \( \pi_0 \) (when \( \mathbf{X} = \mathbf{0} \)) is controlled by the intercept \( \beta_0 \), which is tuned along with \( \beta_{\text{signal}} \) to achieve a specified outcome prevalence and discrimination, key drivers of required sample size in binary prediction models~\cite{Riley2020}. \\\addlinespace[0.75em]
\textbf{Continuous outcomes} &
\( Y \sim \mathcal{N}(\eta, \sigma^2) \), with \( \sigma^2 = 1 \). The signal strength \( \beta_{\text{signal}} \) is tuned to achieve a target large-sample \( R^2 \), a common measure of explained variance in continuous prediction settings~\cite{hawinkel2024out}. \\\addlinespace[0.75em]
\textbf{Survival outcomes} &
Event times \( T \sim \text{Exponential}(\lambda) \), where \( \lambda = \lambda_0 \exp(\eta) \). Right-censoring is introduced at a time point \( t_c \) such that the censoring rate matches a specified value. The linear predictor coefficients are tuned to achieve a target large-sample C-index, the most widely used discrimination measure for time-to-event prediction models~\cite{harrell1996multivariable,uno2011c}.\\
\end{tabular}%
\end{minipage}
\vspace{0.4\baselineskip}

The predictors \( X_j \) are simulated as standardised continuous normal variables, reflecting common design choices in methodological simulation studies \cite{VanCalster2016}. These data-generating processes are implemented internally within the package, with separate routines handling binary, continuous, and survival outcomes, respectively.

\subsubsection{Data generator tuning for continuous outcomes}

The parameters governing each data-generating process are tuned to achieve two user-specified targets: (i) the desired outcome prevalence in the population, and (ii) the maximum achievable performance (e.g., C-statistic for binary/survival, $R^2$ for continuous) of a correctly specified model, as advocated in principled simulation-based evaluations \cite{pavlou2024evaluation,riley2019minimumI}. It should be noted that these tuning strategies are implemented specifically for the default prediction models (logistic, linear, and Cox regression); researchers wishing to employ alternative model classes would need to develop corresponding tuning functions tailored to those models, which remains an avenue for future development.

The continuous outcome tuning function in the package analytically determines the common coefficient \( \beta_{\text{signal}} \) for all signal predictors. Given a target large-sample \( R^2 \), number of candidate features \( p \), and number of noise features \( p_{\text{noise}} \), it calculates
\[
\beta_{\text{signal}} = \sqrt{\frac{R^2}{(p_{\text{signal}} \cdot (1 - R^2))}},
\]
where \( p_{\text{signal}} = p  - p_{\text{noise}} \) is the number of true signal predictors. This derivation follows directly from the relationship between regression coefficients, predictor variance, and explained variance in linear models \cite{kutner2005applied}. This ensures that a linear regression model fitted to a very large sample will achieve the expected \( R^2 \).

\subsubsection{Data generator tuning for binary outcomes}

The tuning of the signal coefficient to match the required performance targets follows an approach similar to that of \cite{Pavlou2021}. A numerical optimisation procedure is employed to jointly calibrate the distribution of the linear predictor \( \eta = \mathbf{X}^T \boldsymbol{\beta} + \beta_0 \), finding the mean \( \mu \) and variance \( \sigma^2 \) of \( \eta \) that simultaneously satisfy the target outcome prevalence \( \pi_0 \) and target large-sample C-statistic. The optimisation minimises the squared error between estimated and target values:
\[
\min_{\mu, \sigma^2} \left[ (\hat{C} - C_{\text{target}})^2 + (\hat{\pi} - \pi_{\text{target}})^2 \right],
\]
where \( \hat{\pi} = \mathbb{E}[\text{logit}^{-1}(\eta)] \) and \( \hat{C} \) is computed via numerical integration of the bivariate normal distribution of the linear predictor for cases and controls, consistent with the binormal model for the ROC curve \cite{Pavlou2021}. Once \( \sigma^2 \) is determined, the common signal coefficient is set as
\[
\beta_{\text{signal}} = \frac{\sigma}{\sqrt{p_{\text{signal}}}},
\]
ensuring that the variance of the linear predictor matches the optimised \( \sigma^2 \).

\subsubsection{Data generator tuning for survival outcomes}
The survival outcome tuning function performs optimisation to find parameters for a proportional hazards model. It searches for the log of the baseline hazard \( \log(\lambda) \) and the log of the standard deviation of the linear predictor \( \log(\sigma) \) that minimises the squared error between the actual event rate (1~$-$~censoring rate) and C-index and their respective targets. The optimisation uses simulated survival data (with exponential event times) to evaluate the objective function, building on established links between the linear predictor variance and concordance in proportional hazards models \cite{Pavlou2021}. The resulting \( \sigma \) is then used to set $\beta_{\text{signal}}$ analogously to the binary case.

\subsection{Model-fitting procedure}
Given a training dataset
\[
\mathcal{D}_{\text{train}} = \{ (\mathbf{X}_i, Y_i) \}_{i=1}^n,
\]
a prediction model \( f(\mathbf{X}; \hat{\boldsymbol{\theta}}) \) is fitted. The package supports logistic regression for binary outcomes, linear regression for continuous outcomes, and Cox proportional hazards models for survival outcomes as default model classes. The model-fitting function \( \mathcal{M} \) returns a fitted model object \( \hat{f} \), which is then used to generate predictions on an independent validation set.

\subsection{Model evaluation and performance metrics}
Model performance is evaluated on a large, independent validation set $\mathcal{D}_{\text{val}}$ of size \( n_{\text{val}}\) to approximate the expected performance in the population. Let \( \hat{\pi}_i \) denote the predicted probability (or risk score) for the \( i \)-th observation. The package supports multiple metrics:

\begin{description}
\item[Discrimination] For binary outcomes, the area under the ROC curve (AUC) is computed. For survival outcomes, Harrell's C-index is used. For continuous outcomes, the out-of-sample \( R^2 \) is calculated as \cite{hawinkel2024out}
\[
R^2 = 1 - \frac{\text{MSE}}{\text{Var}(Y)}.
\]
\item[Calibration] The calibration slope \( \gamma \) is estimated by regressing the observed outcome on the linear predictor (or log-odds) of the predictions. A slope of 1 indicates perfect calibration.
\end{description}

\subsection{Sample size search engines}
In this study, we compare three search engines to find the minimum sample size \( n^* \) such that the expected performance \( \mathbb{E}[G(n)] \) (or a specified quantile thereof) exceeds a target \( \tau \). Let \( \mathcal{S}(n) \) denote the simulation process: generate training data of size \( n \), fit the model, and compute the performance metric \( G \) on the validation set.

\subsubsection{Gaussian process search}\label{sec:gp}
The primary search engine in \texttt{pmsims} is a Gaussian process (GP)-based adaptive procedure implemented via the \texttt{mlpwr} package \cite{zimmer2024sample}. This approach treats the relationship between sample size and model performance as an unknown smooth function,
\[
g(n) = \mathbb{F}\{G(n)\},
\]
and uses GP surrogate modelling to emulate the function using a limited number of simulation evaluations.

\begin{algorithm}[H]
\caption{Gaussian Process--Based Sample Size Search (\texttt{gp} Engine)}\label{algo1}
\begin{algorithmic}[1]
\REQUIRE Data generating function \( D(\cdot) \), model fitting function \( M(\cdot) \), performance metric \( G(\cdot) \)
\REQUIRE Target performance \( \tau \), aggregation criterion (mean or assurance)
\REQUIRE Evaluation budget \( B \), replications per sample size \( \kappa \)
\STATE Compute heuristic starting bounds \( [n_{\min}^{(0)}, n_{\max}^{(0)}] \) based on outcome type and number of predictors \( p \)
\STATE Refine bounds via adaptive search using \( B_0 \) pilot replications, yielding \( [n_{\min}, n_{\max}] \)
\STATE Generate a fixed independent test dataset \( \mathcal{D}_{\text{test}} \leftarrow D(n_{\text{test}}) \)
\STATE Initialise GP with \( n_k \) sample sizes drawn within \( [n_{\min}, n_{\max}] \)
\FOR{each initial sample size \( n_k \)}
  \STATE Simulate training data \( \mathcal{D}_{\text{train}} \leftarrow D(n_k) \)
  \STATE Fit model \( \hat{f} \leftarrow M(\mathcal{D}_{\text{train}}) \)
  \STATE Evaluate performance \( g \leftarrow G(\mathcal{D}_{\text{test}}, \hat{f}) \)
\ENDFOR
\STATE Fit GP surrogate to observed \( (n, g) \) pairs
\WHILE{evaluation budget \( B \) not exhausted}
  \STATE Select next candidate \( n \) via GP acquisition rule, targeting smallest \( n \) achieving \( \tau \)
  \STATE Estimate aggregated performance via \( \kappa \) repeated simulations:
    \IF{criterion is mean}
      \STATE \( \hat{g}(n) \leftarrow \frac{1}{\kappa}\sum_{j=1}^\kappa G(\mathcal{D}_{\text{test}}, M(D_j(n))) \)
    \ELSIF{criterion is assurance}
      \STATE \( \hat{g}(n) \leftarrow Q_{0.20}\bigl(G(\mathcal{D}_{\text{test}}, M(D_j(n)))\bigr) \)
    \ENDIF
  \STATE Update GP surrogate with new \( (n, \hat{g}(n)) \) observation
\ENDWHILE
\RETURN Estimated minimum sample size \( n^* \) such that \( \hat{g}(n^*) \geq \tau \)
\end{algorithmic}
\end{algorithm}

Algorithm~(\ref{algo1}) begins by determining a data-driven lower bound on the sample size using outcome-specific heuristics informed by the number of predictors and, where relevant, the outcome prevalence or censoring rate. A small set of initial sample sizes \( \{n_1, \dots, n_k\} \) is then evaluated to initialise the GP surrogate. At each iteration, the GP guides the selection of a new candidate sample size, targeting the smallest \( n \) at which the desired performance threshold \( \tau \) is achieved.

Model performance is aggregated either by its mean (mean-based criterion) or by a lower quantile of its sampling distribution (assurance criterion). For the assurance criterion, the target is the 20th percentile of the performance distribution across training sets, estimated empirically across simulation replications at each candidate \( n \). In the GP-based engines, simulation noise is explicitly modelled using a bootstrap-based variance estimator, and the search continues until the simulation budget is exhausted.

\subsubsection{Bisection search (\texttt{bisection})}\label{sec:bs}

The bisection search uses a classical deterministic bisection algorithm. This engine maintains an interval \([n_{\text{low}}, n_{\text{high}}]\) that brackets the unknown minimum sample size \( n^* \). At each iteration, the algorithm evaluates model performance at the midpoint
\[
n_{\text{mid}} = \left\lfloor \frac{n_{\text{low}} + n_{\text{high}}}{2} \right\rfloor.
\]
Multiple simulation replicates are generated at \( n_{\text{mid}} \), and performance is aggregated using either the mean-based or assurance criterion. If the aggregated performance estimate \( \hat{g}(n_{\text{mid}}) \ge \tau \), the upper bound is updated to \( n_{\text{mid}} \); otherwise, the lower bound is updated. The interval is repeatedly halved until its width falls below a predefined tolerance or the simulation budget is exhausted. 

\begin{algorithm}[H]
\caption{Bisection-Based Sample Size Search (\texttt{bisection} Engine)}\label{algo2}
\begin{algorithmic}[1]
\REQUIRE Data generating function \( D(\cdot) \), model fitting function \( M(\cdot) \), performance metric \( G(\cdot) \)
\REQUIRE Target performance \( \tau \), aggregation criterion (mean or assurance)
\REQUIRE Evaluation budget \( B \), replications per sample size \( \kappa \)
\STATE Compute heuristic starting bounds \( [n_{\min}^{(0)}, n_{\max}^{(0)}] \) based on outcome type and number of predictors \( p \)
\STATE Refine bounds via adaptive search using \( B_0 \) pilot replications, yielding \( [n_{\min}, n_{\max}] \)
\STATE Generate a fixed independent test dataset \( \mathcal{D}_{\text{test}} \leftarrow D(n_{\text{test}}) \)
\STATE Set maximum iterations \( T \leftarrow \lfloor B / \kappa \rfloor \)
\STATE Evaluate aggregated performance at bounds: \( \hat{g}(n_{\min}) \) and \( \hat{g}(n_{\max}) \)
\WHILE{iteration \( t < T \)}
  \STATE Set \( n_{\text{mid}} \leftarrow \left\lfloor (n_{\min} + n_{\max}) / 2 \right\rfloor \)
  \STATE Simulate \( \kappa \) training datasets \( \mathcal{D}_{\text{train}} \leftarrow D_j(n_{\text{mid}}),\ j = 1, \dots, \kappa \)
  \STATE Fit model \( \hat{f}_j \leftarrow M(\mathcal{D}_{\text{train},j}) \) and evaluate \( G(\mathcal{D}_{\text{test}}, \hat{f}_j) \) for each replicate
  \STATE Aggregate performance:
    \IF{criterion is mean}
      \STATE \( \hat{g}(n_{\text{mid}}) \leftarrow \frac{1}{\kappa}\sum_{j=1}^\kappa G(\mathcal{D}_{\text{test}}, \hat{f}_j) \)
    \ELSIF{criterion is assurance}
      \STATE \( \hat{g}(n_{\text{mid}}) \leftarrow Q_{0.20}\bigl(G(\mathcal{D}_{\text{test}}, \hat{f}_j)\bigr) \)
    \ENDIF
  \IF{\( \hat{g}(n_{\text{mid}}) \geq \tau \)}
    \STATE \( n_{\max} \leftarrow n_{\text{mid}} \),\quad \( \hat{g}(n_{\max}) \leftarrow \hat{g}(n_{\text{mid}}) \)
  \ELSE
    \STATE \( n_{\min} \leftarrow n_{\text{mid}} \),\quad \( \hat{g}(n_{\min}) \leftarrow \hat{g}(n_{\text{mid}}) \)
  \ENDIF
  \STATE \( t \leftarrow t + 1 \)
\ENDWHILE
\RETURN Estimated minimum sample size \( n^* = n_{\max} \)
\end{algorithmic}
\end{algorithm}

\subsubsection{Hybrid GP--bisection search (\texttt{gp-bs})}\label{sec:gpbs}

The hybrid engine (\texttt{gp\_bs}) combines bisection and Gaussian process surrogate modelling in a three-stage strategy. First, adaptive bound estimation is performed as in the other engines. Second, a coarse bisection search is run with a fixed budget of \( T = \lfloor0.2 \times B\rfloor \) replications to rapidly narrow the search interval to \( [n_{\min}^*, n_{\max}^*] \), reducing the risk of poor GP initialisation that can occur when the search space is wide or the performance function is highly variable. The refined interval is then passed to the GP search stage, which conducts an adaptive, model-guided search within these tighter bounds. This engine may be particularly useful in scenarios with high performance variance or complex data-generating mechanisms.

\begin{algorithm}[H]
\caption{Hybrid Bisection--Gaussian Process Search (\texttt{gp-bs} Engine)}
\begin{algorithmic}[1]
\REQUIRE Data generating function \( D(\cdot) \), model fitting function \( M(\cdot) \), performance metric \( G(\cdot) \)
\REQUIRE Target performance \( \tau \), aggregation criterion (mean or assurance)
\REQUIRE Evaluation budget \( B \), replications per sample size \( \kappa \)

\textit{\% Stage 1: Adaptive bound estimation}
\STATE Compute heuristic starting bounds \( [n_{\min}^{(0)}, n_{\max}^{(0)}] \) based on outcome type and number of predictors \( p \)
\STATE Refine bounds via adaptive search using \( B_0 \) pilot replications, yielding \( [n_{\min}, n_{\max}] \)
\STATE Generate a fixed independent test dataset \( \mathcal{D}_{\text{test}} \leftarrow D(n_{\text{test}}) \)

\textit{\% Stage 2: Coarse bisection search}
\STATE Run bisection algorithm~(\ref{algo2}) within \( [n_{\min}, n_{\max}] \) using a fixed budget of \( T = \lfloor0.2 \times B\rfloor \) total replications, yielding bisection history \( \mathcal{H} = \{(n_t, \hat{g}(n_t))\}_{t=1}^T \)

\textit{\% Stage 3: Refined GP search}
\STATE Derive refined GP bounds \( [n_{\min}^*, n_{\max}^*] \) from \( \mathcal{H} \)
\STATE Initialise GP with \( k \) start sets drawn within \( [n_{\min}^*, n_{\max}^*] \)
\WHILE{evaluation budget \( B \) not exhausted}
  \STATE Select next candidate \( n \) via GP acquisition rule, targeting smallest \( n \) achieving \( \tau \)
  \STATE Simulate \( \kappa \) training datasets \( \mathcal{D}_{\text{train}} \leftarrow D_j(n),\ j = 1, \dots, \kappa \)
  \STATE Fit model \( \hat{f}_j \leftarrow M(\mathcal{D}_{\text{train},j}) \) and evaluate \( G(\mathcal{D}_{\text{test}}, \hat{f}_j) \) for each replicate
  \STATE Aggregate performance:
    \IF{criterion is mean}
      \STATE \( \hat{g}(n) \leftarrow \frac{1}{\kappa}\sum_{j=1}^\kappa G(\mathcal{D}_{\text{test}}, \hat{f}_j) \)
    \ELSIF{criterion is assurance}
      \STATE \( \hat{g}(n) \leftarrow Q_{0.20}\bigl(G(\mathcal{D}_{\text{test}}, \hat{f}_j)\bigr) \)
    \ENDIF
  \STATE Update GP surrogate with new \( (n, \hat{g}(n)) \) observation
\ENDWHILE
\RETURN Estimated minimum sample size \( n^* \) such that \( \hat{g}(n^*) \geq \tau \)
\end{algorithmic}
\end{algorithm}

\subsection{Adaptive starting value search for \texttt{pmsims} engines}\label{startvalues}

All three engines share a common first stage: an adaptive procedure that automatically determines the sample size bounds \( [n_{\min}, n_{\max}] \) supplied to the main search algorithm (as described in Sections~\ref{sec:gp}--\ref{sec:gpbs}). The goal is to find two sample sizes \( n_{\min} \) and \( n_{\max} \) such that
\[
\hat{G}(n_{\min}) \leq \tau \leq \hat{G}(n_{\max}),
\]
where \( \tau \) is the user-specified target performance. This stage uses a fixed pilot budget of $B_0$ replications (distinct from the main evaluation budget \( B \)), allocated as \( \kappa = \lfloor B_0 / K \rfloor \) replications per candidate sample size, where \( K \) is the maximum number of iterations. A tolerance \( \delta = 0.0001 \) is applied to avoid unnecessary iterations when the performance estimate is already sufficiently close to \( \tau \).

\subsubsection{Initialisation}

\begin{enumerate}
\item Generate a fixed test set \( \mathcal{D}_{\text{test}} \leftarrow D(n_{\text{test}}) \).
\item Set \( k \leftarrow 1 \), \( n^{(1)} \leftarrow n_0 \), where \( n_0 \) is a heuristic starting value based on outcome type and number of predictors \( p \).
\item Compute \( \hat{G}^{(1)} \) using \( \kappa \) replications.
\item Determine the direction of search:
\[
\text{direction} \leftarrow \begin{cases}
\text{``up''} & \text{if } \hat{G}^{(1)} < \tau,\\
\text{``down''} & \text{if } \hat{G}^{(1)} \ge \tau.
\end{cases}
\]
\end{enumerate}

\subsubsection{Iterative search}

While \( k < K \):

\begin{enumerate}
\item Propose a new candidate sample size:
\[
n^{(k+1)} \leftarrow \begin{cases}
2\, n^{(k)} & \text{if direction = ``up'',}\\
\lfloor n^{(k)} / 2 \rfloor & \text{if direction = ``down''.}
\end{cases}
\]
If \( n^{(k+1)} = n^{(k)} \), the algorithm terminates.

\item Compute \( \hat{G}^{(k+1)} \) using \( \kappa \) fresh replications.

\item Update bounds:
\begin{itemize}
\item[-] If direction = ``up'':
  \begin{itemize}
  \item[-] If \( \hat{G}^{(k+1)} \ge \tau - \delta \): set \( n_{\max} \leftarrow n^{(k+1)} \) and stop.
  \item[-] Otherwise: set \( n_{\min} \leftarrow n^{(k+1)} \) and continue.
  \end{itemize}
\item[-] If direction = ``down'':
  \begin{itemize}
  \item[-] If \( \hat{G}^{(k+1)} \le \tau + \delta \): set \( n_{\min} \leftarrow n^{(k+1)} \) and stop.
  \item[-] Otherwise: set \( n_{\max} \leftarrow n^{(k+1)} \) and continue.
  \end{itemize}
\end{itemize}
Set \( k \leftarrow k + 1 \) and repeat.
\end{enumerate}

The doubling and halving strategy locates a bracket around \( \tau \) without requiring prior knowledge of the sample size--performance relationship. The procedure is intentionally conservative, using few replications per candidate to keep the pilot cost low. The method assumes that the performance metric is monotonically non-decreasing in \( n \); if the metric is non-monotonic, the algorithm may fail to bracket the target correctly, and manual specification of \( [n_{\min}, n_{\max}] \) is recommended.

\section{Simulation scenarios for \texttt{pmsims} package validation}

This simulation study had two aims, following standard operating simulation design principles \cite{Moons2015}. Aim~1 evaluated the statistical performance and computational efficiency of the Gaussian process--based adaptive search \texttt{gp} engine, benchmarking it against two reference engines: its hybrid (\texttt{gp-bs}) and a bisection search (\texttt{bisection}). Aim~2 benchmarked \texttt{pmsims}, using the engine identified as best-performing in Aim~1, against two existing approaches: \texttt{pmsampsize} \cite{Riley2020} and \texttt{samplesizedev} \cite{pavlou2024evaluation}.

\subsection{Aim~1: Comparison of \texttt{pmsims} search engines}

To systematically compare the performance of the three search engines (\texttt{gp}, \texttt{bisection}, and \texttt{gp-bs}), we conducted an extensive simulation study covering a range of realistic prediction modelling scenarios, consistent with best-practice guidance for simulation-based methodological research \cite{Burton2006,Morris2019}. The scenarios were varied along five key dimensions: (1) outcome type, (2) number of candidate predictors, (3) outcome distribution or model strength, (4) target performance metric and threshold, and (5) total simulation budget. For each outcome type---binary, continuous, and survival---we defined a set of base scenarios reflecting common characteristics encountered in clinical prediction research \cite{Steyerberg2010,Riley2020}. Each scenario was independently replicated 100 times to ensure robust estimation of the precision and stability of the sample size estimates produced by each engine, in line with recommendations for Monte Carlo precision assessment \cite{Morris2019}.

The levels of specific factors for each outcome type are summarised in Table~\ref{tab:simulation_scenarios}. For binary outcomes, we considered two event prevalences (5\% and 20\%) and two target metrics: the C-statistic (AUC) with a target set 0.05 below the large-sample value and the calibration slope with a target of 0.9. For continuous outcomes, we varied the expected large-sample \( R^2 \) (0.2 and 0.7) and similarly targeted either \( R^2 \) or a calibration slope of 0.9. For survival outcomes, we considered two event rates (40\% and 80\%) with targets on the C-index or the calibration slope (0.9).

Across all outcome types, we examined both a modest (10) and a larger (100) number of candidate predictors, with no noise predictors. The total simulation budget \( B \) was varied from \(200\) to \(2{,}000\) to assess how algorithmic efficiency scales with available computational resources. The budget was allocated in blocks of \( \kappa \in \{10, 20\} \) independent simulation replicates per candidate sample size.

\begin{table}[ht]
\centering
\caption{Simulation scenarios for the comparative evaluation of search engines.}
\label{tab:simulation_scenarios}
\small
\begin{tabular}{lllcc}
\toprule
\textbf{Outcome} & \textbf{\(p\)} & \textbf{Model strength} & \textbf{Target metric} & \textbf{Target value} \\
\midrule
\multirow{4}{*}{Binary}     & \multirow{4}{*}{10, 100} & Prevalence = 0.05, \(C = 0.80\) & AUC               & 0.75 \\
                           &                          & Prevalence = 0.05, \(C = 0.80\) & Calibration slope & 0.90 \\
                           &                          & Prevalence = 0.20, \(C = 0.80\) & AUC               & 0.75 \\
                           &                          & Prevalence = 0.20, \(C = 0.80\) & Calibration slope & 0.90 \\
\midrule
\multirow{4}{*}{Continuous} & \multirow{4}{*}{10, 100} & \(R^2 = 0.20\)                  & \(R^2\)           & 0.15 \\
                            &                          & \(R^2 = 0.20\)                  & Calibration slope & 0.90 \\
                            &                          & \(R^2 = 0.70\)                  & \(R^2\)           & 0.65 \\
                            &                          & \(R^2 = 0.70\)                  & Calibration slope & 0.90 \\
\midrule
\multirow{4}{*}{Survival}   & \multirow{4}{*}{10, 100} & Event rate = 0.40, \(C = 0.80\) & C-index           & 0.75 \\
                            &                          & Event rate = 0.40, \(C = 0.80\) & Calibration slope & 0.90 \\
                            &                          & Event rate = 0.80, \(C = 0.80\) & C-index           & 0.75 \\
                            &                          & Event rate = 0.80, \(C = 0.80\) & Calibration slope & 0.90 \\
\bottomrule
\end{tabular}
\end{table}

\subsection{Aim~2: Benchmark comparison of \texttt{pmsims} with \texttt{pmsampsize} and \texttt{samplesizedev}}

To benchmark the simulation-based approach implemented in \texttt{pmsims} against existing methods, we compared the sample size estimates from the best-performing engine (\texttt{gp}) with those provided by \texttt{pmsampsize} (for binary and continuous outcomes) and \texttt{samplesizedev} (for binary and survival outcomes). We evaluated a range of realistic prediction-modelling scenarios, systematically varying the model strength, outcome prevalence (or event rate), number of predictors, and target performance metric.

For binary outcomes, we considered large-sample C-statistics (AUC) of 0.8 and 0.9; outcome prevalences of 5\% and 20\%; and predictor counts ranging from 5 to 100 in seven steps (5, 10, 20, 40, 60, 80, 100). For continuous outcomes, the large-sample \(R^2\) was set at 0.2, 0.5, and 0.7, with the same range of predictor counts. For survival outcomes, we used large-sample C-indices of 0.7, 0.8, and 0.9; event rates of 40\%, 50\%, and 80\%; and the same predictor counts.

Within each scenario, two target metrics were examined: a discrimination metric set to 0.05 below the large-sample value, and the calibration slope set to 0.90. The \texttt{gp} engine was run with a fixed simulation budget of 1{,}000 total model evaluations, using 20 replications per candidate sample size, under both mean-based and assurance-based criteria (80\% assurance). For each scenario, we recorded the recommended sample size for each method and the corresponding computational time. To assess validity, we generated an independent validation dataset of 30{,}000 observations and evaluated the achieved performance of a model developed on a sample of the recommended size. Each scenario was repeated 100 times. The complete set of evaluated scenarios is summarised in Table~\ref{tab:analytical_scenarios}.

\begin{table}[ht]
\centering
\caption{Simulation scenarios for comparing \texttt{pmsims} (\texttt{gp} engine) with existing sample size tools (\texttt{pmsampsize}, \texttt{samplesizedev}). All scenarios were evaluated for both mean-based and assurance-based criteria (80\% assurance).}
\label{tab:analytical_scenarios}
\begin{tabular}{llclcc}
\toprule
\shortstack[l]{Outcome\\type} &
\shortstack[l]{Large-sample\\performance} &
\shortstack[l]{Prevalence / event rate} &
\shortstack[l]{Target\\metric} &
\shortstack[l]{Target\\value} \\
\midrule
Binary 
& C-statistic $= 0.8, 0.9$ 
& 0.05, 0.20 
& AUC 
& $|C - 0.05|$ \\
& C-statistic $= 0.8, 0.9$ 
& 0.05, 0.20 
& Calibration slope 
& 0.90 \\
Continuous 
& $R^2 = 0.2, 0.5, 0.7$ 
& -- 
& $R^2$ 
& $|R^2 - 0.05|$ \\
& $R^2 = 0.2, 0.5, 0.7$ 
& -- 
& Calibration slope 
& 0.90 \\
Survival 
& C-index $= 0.7, 0.8, 0.9$ 
& 0.40, 0.50, 0.80 
& C-index 
& $|C - 0.05|$ \\
& C-index $= 0.7, 0.8, 0.9$ 
& 0.40, 0.50, 0.80 
& Calibration slope 
& 0.90 \\
\bottomrule
\end{tabular}
\end{table}

\noindent\textit{Note:} For binary and survival outcomes, the large-sample performance is the expected C-statistic (AUC) or C-index, respectively. The target for discrimination metrics is set 0.05 below this value to reflect a practically acceptable margin for finite-sample optimism \cite{Steyerberg2010}. The calibration-slope target is 0.90, representing a commonly accepted threshold for good calibration in clinical prediction models \cite{VanCalster2016}. All predictors are continuous and truly associated with the outcome.

\subsection{Simulation estimands}

In each simulation scenario, the performance of the sample size determination procedures is summarised using Monte Carlo estimands computed across \( S = 100 \) independent simulation runs \cite{Morris2019, Burton2006}. We computed the mean \( \hat{n}^* \) and standard deviation \( \sigma_{\hat{n}^*} \) of the estimated minimum sample sizes across runs, as well as the coefficient of variation,
\[
\text{CV} = \frac{\sigma_{\hat{n}^*}}{\hat{n}^*} \times 100\%,
\]
which provides a scale-free measure of relative dispersion and is used to assess estimator stability across simulation replicates \cite{Burton2006}.

For Aim~2, performance is evaluated based on the \emph{achieved model performance} when fitting a model using the sample size recommended by each method. Let \( \widehat{\text{Perf}}^{(s)} \) denote the performance estimated on an independent validation dataset in replicate \( s \), and let \( \text{Target} \) denote the corresponding target performance value. Performance deviation is defined as
\[
\text{Deviation}^{(s)}(\%) =
\frac{\widehat{\text{Perf}}^{(s)} - \text{Target}}
     {\text{Target}} \times 100.
\]

\section{Results}
 
\subsection{Aim~1: Comparison of search engines}
 
\subsubsection*{Binary outcomes}
 
\begin{figure}[H]
    \centering
    \begin{subfigure}[t]{0.97\linewidth}
        \centering
        \includegraphics[width=\linewidth]{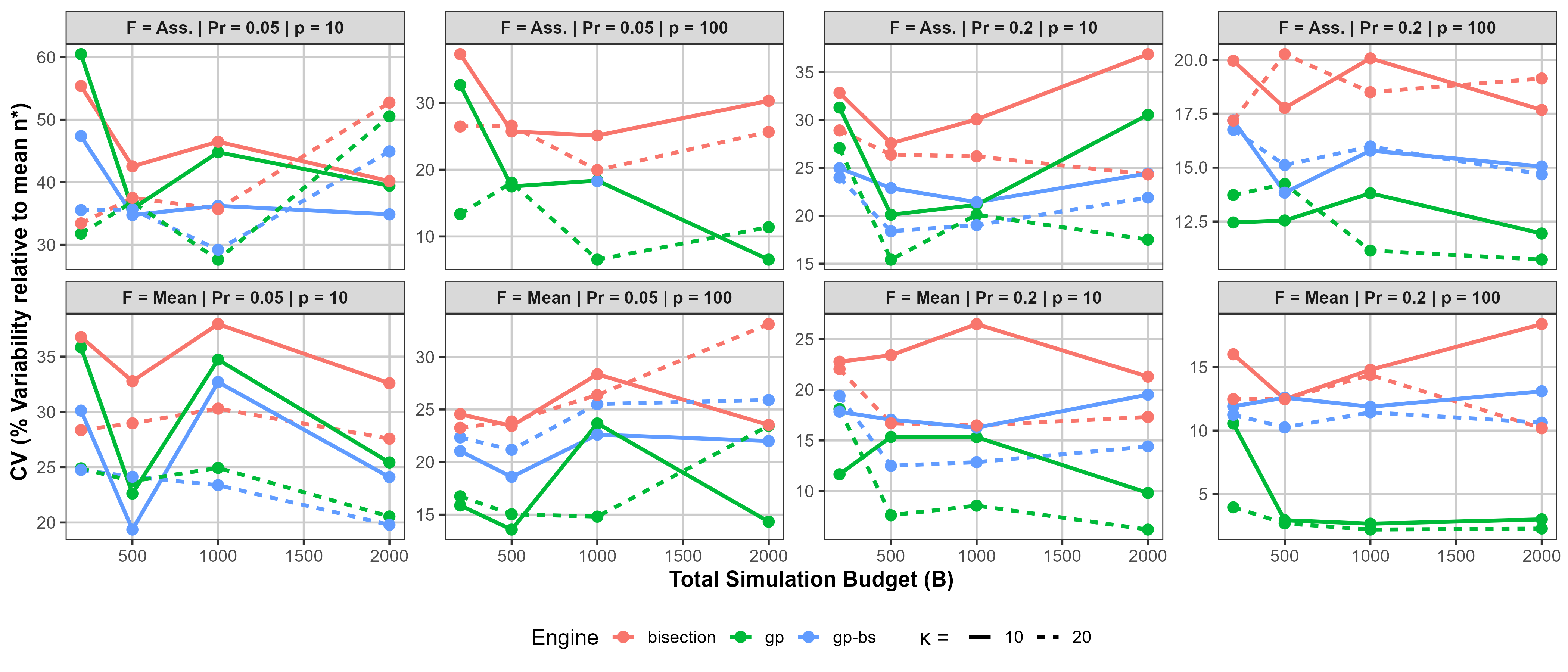}
        \caption{Coefficient of Variation (CV) of sample size estimates relative to mean $n^*$.}
        \label{fig:aim1-cv}
    \end{subfigure}
    \hspace{0.5cm}
    \begin{subfigure}[t]{0.97\linewidth}
        \centering
        \includegraphics[width=\linewidth]{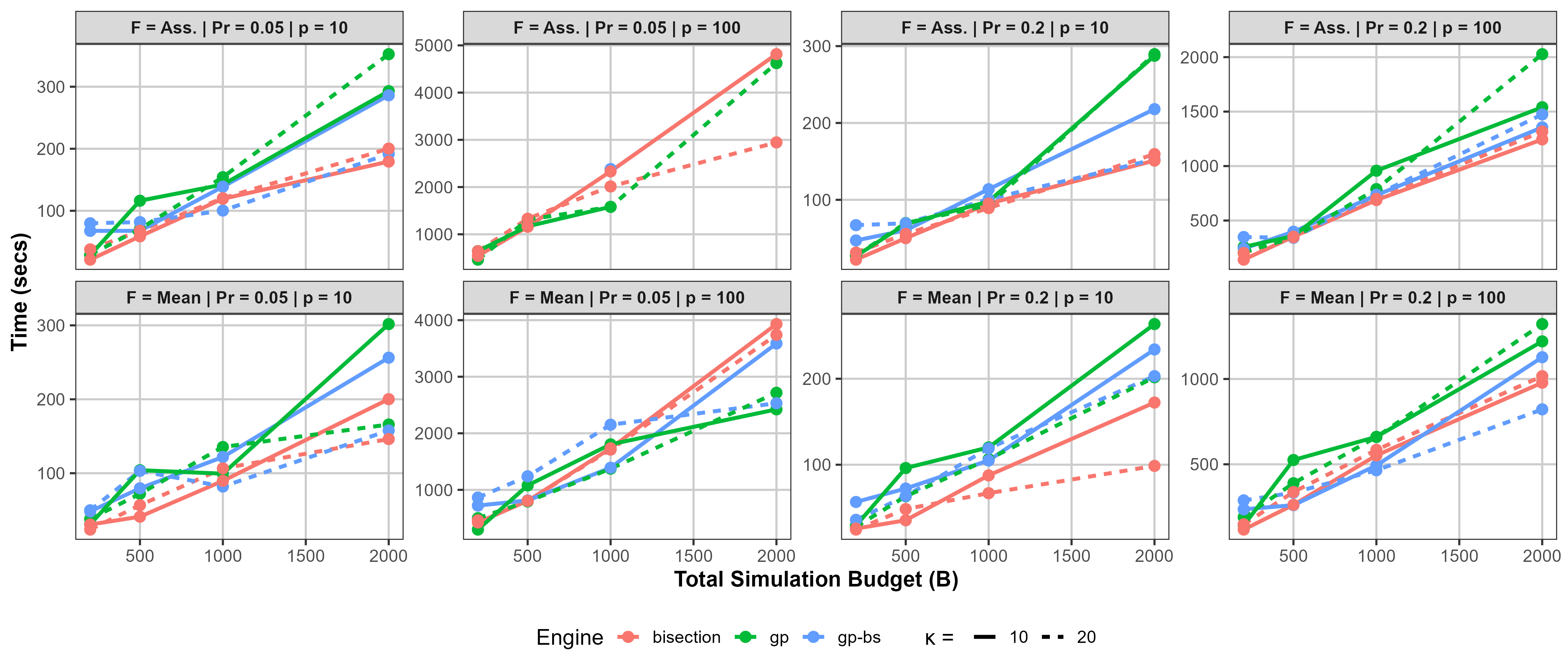}
        \caption{Computational time required to estimate minimum sample size $n^*$.}
        \label{fig:aim1-time}
    \end{subfigure}
    \caption{Aim~1 (Binary outcome): Comparison of CV and computational time across search engines \texttt{gp}, \texttt{bisection} and \texttt{gp-bs} for calibration slope metric under varying number of simulation replicates per evaluation $(\kappa = 10, 20)$, prevalence $(0.05, 0.20)$, number of predictors $(p = 10, 100)$ and total simulation budget $(B = 200$--$2000)$.}
    \label{fig:aim1-binary-summary}
\end{figure}

Figure~\ref{fig:aim1-binary-summary} compares precision (coefficient of variation; CV) and computational time for the three search procedures across the calibration slope metric under varying design parameters: number of simulation replicates per evaluation ($\kappa = 10, 20$), outcome prevalence ($0.05, 0.20$), number of predictors ($p = 10, 100$), and total simulation budget ($B = 200$--$2000$). Increasing the number of replicates per evaluation to $20$ substantially reduced the CV for the \texttt{gp} engine. Across all remaining conditions, \texttt{gp} consistently produced lower and more stable CV estimates than \texttt{bisection} and \texttt{gp-bs}, with the advantage most pronounced in the low-prevalence setting. Performance improvements were more consistent under mean-aggregation, which yielded smoother convergence trajectories. On average, with $\kappa = 20$, CV plateaued at a total simulation budget of approximately $B = 1000$ under both aggregation schemes, indicating that $B = 1000$ constitutes an efficient lower bound. Comparable patterns were obtained for the AUC metric (Figure~\ref{fig:aim1-binary-summary2}, Additional~File~1).
 
At $B = 1000$ (Table~\ref{tab:binary_combined}), \texttt{gp} consistently achieved the lowest CV for AUC estimation, most notably in challenging settings characterised by low prevalence ($0.05$) and high dimensionality ($p = 100$). For calibration slope estimation, \texttt{gp-bs} occasionally outperformed the other methods under assurance-based aggregation in low-prevalence scenarios, though \texttt{gp} generally delivered competitive performance under mean aggregation.
 
\begin{table}[H]
\centering
\caption{\label{tab:binary_combined}Estimated minimum sample size ($\hat{n}^*$) and stability (CV) for binary outcomes using assurance aggregation ($20\%$ quantile) and mean aggregation, stratified by target metric, event prevalence and number of predictors ($p$), at a fixed total budget $B = 1000$ and $\kappa = 20$. Results are averaged over 100 simulation replicates.}
\centering
\fontsize{8}{9}\selectfont
\begin{tabular}[t]{ccclcccc}
\toprule
 \textbf{Metric} & \textbf{Prevalence} & $\bm p$ & \textbf{Engine} & \multicolumn{2}{c}{\textbf{Mean} $\hat{n}^*$} & \multicolumn{2}{c}{\textbf{CV}} \\
 &  &  &  & Assurance & Mean agg. & Assurance & Mean agg. \\
\midrule
\multirow{24}{*}[3.5\dimexpr\aboverulesep+\belowrulesep+\cmidrulewidth]{\raggedleft\arraybackslash }
   &  &  & bisection & 444 & 320 & 17.05 & 11.79\\
   &  &  & \textbf{gp} & \textbf{431} & \textbf{315} & \textbf{13.39} & \textbf{3.01}\\
   &  & \multirow{-3}{*}{\raggedright\arraybackslash 10} & gp-bs & 428 & 308 & 11.72 & 10.11\\
\cmidrule{3-8}
   &  &  & bisection & 3671 & 3190 & 11.67 & 8.17\\
   &  &  & \textbf{gp} & \textbf{3691} & \textbf{3168} & \textbf{5.43} & \textbf{1.52}\\
   & \multirow{-6}{*}[0.5\dimexpr\aboverulesep+\belowrulesep+\cmidrulewidth]{\raggedright\arraybackslash 0.05} & \multirow{-3}{*}{\raggedright\arraybackslash 100} & gp-bs & 3650 & 3127 & 10.08 & 9.90\\
\cmidrule{2-8}
   &  &  & bisection & 152 & 111 & 12.68 & 10.89\\
   &  &  & \textbf{gp} & \textbf{154} & \textbf{110} & \textbf{4.51} & \textbf{2.91}\\
   &  & \multirow{-3}{*}{\raggedright\arraybackslash 10} & gp-bs & 152 & 110 & 6.17 & 7.03\\
\cmidrule{3-8}
   &  &  & bisection & 1311 & 1145 & 7.43 & 6.76\\
   &  &  & \textbf{gp} & \textbf{1307} & \textbf{1138} & \textbf{2.61} & \textbf{0.82}\\
  \multirow{-12}{*}[1.5\dimexpr\aboverulesep+\belowrulesep+\cmidrulewidth]{\raggedright\arraybackslash AUC} & \multirow{-6}{*}[0.5\dimexpr\aboverulesep+\belowrulesep+\cmidrulewidth]{\raggedright\arraybackslash 0.2} & \multirow{-3}{*}{\raggedright\arraybackslash 100} & gp-bs & 1289 & 1145 & 6.09 & 6.35\\
\midrule[\heavyrulewidth]
   &  &  & bisection & 3778 & 1641 & 35.72 & 30.30\\
   &  &  & gp & 3515 & 1566 & 27.62 & 24.94\\
   &  & \multirow{-3}{*}{\raggedright\arraybackslash 10} & \textbf{gp-bs} & \textbf{3577} & \textbf{1596} & \textbf{29.20} & \textbf{23.37}\\
\cmidrule{3-8}
   &  &  & bisection & 23654 & 18314 & 19.94 & 26.38\\
   &  &  & \textbf{gp} & \textbf{23955} & \textbf{17958} & \textbf{6.53} & \textbf{14.82}\\
   & \multirow{-6}{*}[0.5\dimexpr\aboverulesep+\belowrulesep+\cmidrulewidth]{\raggedright\arraybackslash 0.05} & \multirow{-3}{*}{\raggedright\arraybackslash 100} & gp-bs & 22669 & 18879 & 18.28 & 25.52\\
\cmidrule{2-8}
   &  &  & bisection & 1389 & 594 & 26.20 & 16.48\\
   &  &  & gp & 1311 & 592 & 20.11 & 8.57\\
   &  & \multirow{-3}{*}{\raggedright\arraybackslash 10} & \textbf{gp-bs} & \textbf{1325} & \textbf{583} & \textbf{19.03} & \textbf{12.86}\\
\cmidrule{3-8}
   &  &  & bisection & 8704 & 6696 & 18.50 & 14.39\\
   &  &  & \textbf{gp} & \textbf{8654} & \textbf{6432} & \textbf{11.15} & \textbf{2.19}\\
  \multirow{-12}{*}[1.5\dimexpr\aboverulesep+\belowrulesep+\cmidrulewidth]{\raggedright\arraybackslash Calib. Slope} & \multirow{-6}{*}[0.5\dimexpr\aboverulesep+\belowrulesep+\cmidrulewidth]{\raggedright\arraybackslash 0.2} & \multirow{-3}{*}{\raggedright\arraybackslash 100} & gp-bs & 8308 & 6595 & 15.98 & 11.44\\
\bottomrule
\end{tabular}
\end{table}

\subsubsection*{Continuous outcomes}
 
Results for continuous outcomes were consistent with the binary findings. The \texttt{gp} engine consistently produced the lowest and most stable CV across all experimental conditions, with the advantage particularly pronounced in high-dimensional settings ($p = 100$) and under low-signal conditions ($R^2 = 0.2$), where \texttt{gp} yielded CV reductions of up to 70--80\% relative to \texttt{bisection} and \texttt{gp-bs}. With $\kappa = 20$, CV plateaued at approximately $B = 1000$. Figures~\ref{fig:aim1-continuous-summary} and~\ref{fig:aim1-continuous-summary2} (Additional~File~1) display CV and computational time for the calibration slope and $R^2$ metrics respectively, and the full numerical results are reported in Table~\ref{tab:continuous_combined}.
 
\begin{table}[H]
\centering
\caption{\label{tab:continuous_combined}
Estimated minimum sample size ($\hat{n}^*$) and stability (CV) for continuous outcomes using assurance aggregation ($20\%$ quantile) and mean aggregation, stratified by target metric, large-sample $R^2$, and number of predictors ($p$), at $B = 1000$ and $\kappa = 20$. Results are averaged over 100 simulation replicates.}
\centering
\fontsize{8}{9}\selectfont
\begin{tabular}[t]{llllcccc}
\toprule
\textbf{Metric} & $\bm{R^2}$ & $\bm p$ & \textbf{Engine} & \multicolumn{2}{c}{\textbf{Mean} $\hat{n}^*$} & \multicolumn{2}{c}{\textbf{CV}} \\
 &  &  &  & Assurance & Mean agg. & Assurance & Mean agg. \\
\midrule
\multirow{12}{*}[1.5\dimexpr\aboverulesep+\belowrulesep+\cmidrulewidth]{\raggedright\arraybackslash $R^2$}
 &  &  & bisection & 243 & 191 & 11.20 & 10.46 \\
 &  &  & \textbf{gp} & \textbf{237} & \textbf{191} & \textbf{6.65} & \textbf{3.62} \\
 &  & \multirow{-3}{*}{\raggedright\arraybackslash 10} & gp-bs & 242 & 188 & 8.51 & 6.71 \\
\cmidrule{3-8}
 &  &  & bisection & 1934 & 1741 & 7.04 & 8.33 \\
 &  &  & \textbf{gp} & \textbf{1954} & \textbf{1721} & \textbf{1.85} & \textbf{3.46} \\
 & \multirow{-6}{*}[0.5\dimexpr\aboverulesep+\belowrulesep+\cmidrulewidth]{\raggedright\arraybackslash 0.2}
   & \multirow{-3}{*}{\raggedright\arraybackslash 100} & gp-bs & 1907 & 1735 & 7.97 & 7.18 \\
\cmidrule{2-8}
 &  &  & bisection & 99 & 77 & 8.56 & 11.31 \\
 &  &  & \textbf{gp} & \textbf{100} & \textbf{79} & \textbf{2.52} & \textbf{6.22} \\
 &  & \multirow{-3}{*}{\raggedright\arraybackslash 10} & gp-bs & 99 & 78 & 6.17 & 6.37 \\
\cmidrule{3-8}
 &  &  & bisection & 786 & 710 & 6.30 & 6.26 \\
 &  &  & \textbf{gp} & \textbf{785} & \textbf{707} & \textbf{0.86} & \textbf{0.61} \\
 & \multirow{-6}{*}[0.5\dimexpr\aboverulesep+\belowrulesep+\cmidrulewidth]{\raggedright\arraybackslash 0.7}
   & \multirow{-3}{*}{\raggedright\arraybackslash 100} & gp-bs & 778 & 711 & 5.72 & 5.08 \\
\midrule[\heavyrulewidth]
\multirow{12}{*}[1.5\dimexpr\aboverulesep+\belowrulesep+\cmidrulewidth]{\raggedright\arraybackslash Calib.\ Slope}
 &  &  & bisection & 86 & 190 & 17.56 & 8.03 \\
 &  &  & \textbf{gp} & \textbf{91} & \textbf{191} & \textbf{10.21} & \textbf{3.55} \\
 &  & \multirow{-3}{*}{\raggedright\arraybackslash 10} & gp-bs & 84 & 187 & 12.02 & 7.01 \\
\cmidrule{3-8}
 &  &  & bisection & 4676 & 1741 & 12.59 & 8.50 \\
 &  &  & \textbf{gp} & \textbf{4530} & \textbf{1732} & \textbf{6.30} & \textbf{4.34} \\
 & \multirow{-6}{*}[0.5\dimexpr\aboverulesep+\belowrulesep+\cmidrulewidth]{\raggedright\arraybackslash 0.2}
   & \multirow{-3}{*}{\raggedright\arraybackslash 100} & gp-bs & 4594 & 1745 & 12.54 & 8.46 \\
\cmidrule{2-8}
 &  &  & bisection & 598 & 77 & 7.51 & 7.71 \\
 &  &  & \textbf{gp} & \textbf{582} & \textbf{78} & \textbf{3.71} & \textbf{1.81} \\
 &  & \multirow{-3}{*}{\raggedright\arraybackslash 10} & gp-bs & 580 & 77 & 5.13 & 5.28 \\
\cmidrule{3-8}
 &  &  & bisection & 572 & 715 & 8.56 & 5.94 \\
 &  &  & \textbf{gp} & \textbf{568} & \textbf{706} & \textbf{5.35} & \textbf{0.49} \\
 & \multirow{-6}{*}[0.5\dimexpr\aboverulesep+\belowrulesep+\cmidrulewidth]{\raggedright\arraybackslash 0.7}
   & \multirow{-3}{*}{\raggedright\arraybackslash 100} & gp-bs & 583 & 714 & 4.33 & 5.50 \\
\bottomrule
\end{tabular}
\end{table}

\subsubsection*{Survival outcomes}
 
\begin{figure}[H]
    \centering
    \begin{subfigure}[t]{0.95\linewidth}
        \centering
        \includegraphics[width=\linewidth]{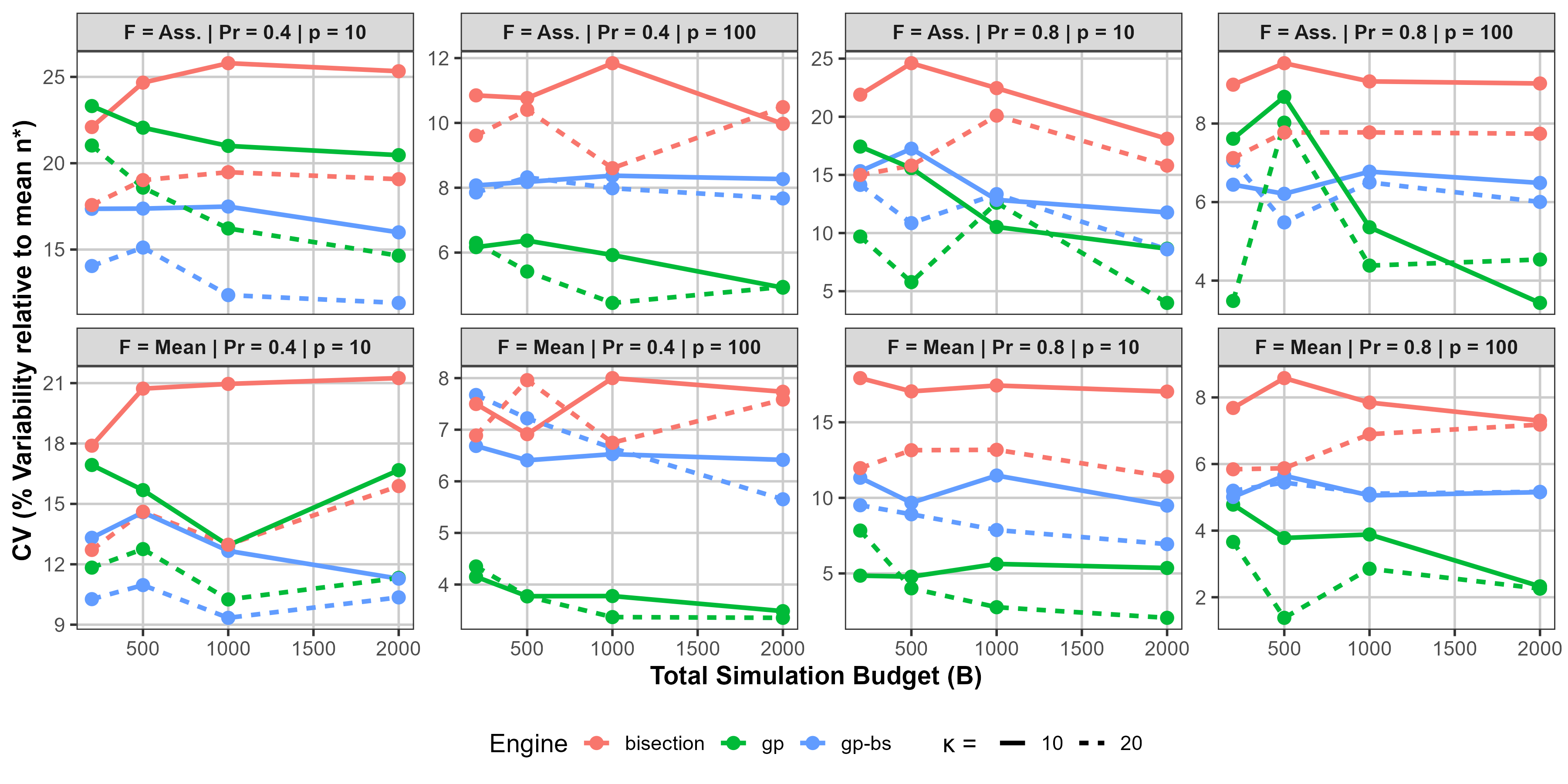}
        \caption{Coefficient of Variation (CV) of sample size estimates relative to mean $n^*$.}
        \label{fig:aim1-surv-cv}
    \end{subfigure}
    \hspace{0.5cm}
    \begin{subfigure}[t]{0.95\linewidth}
        \centering
        \includegraphics[width=\linewidth]{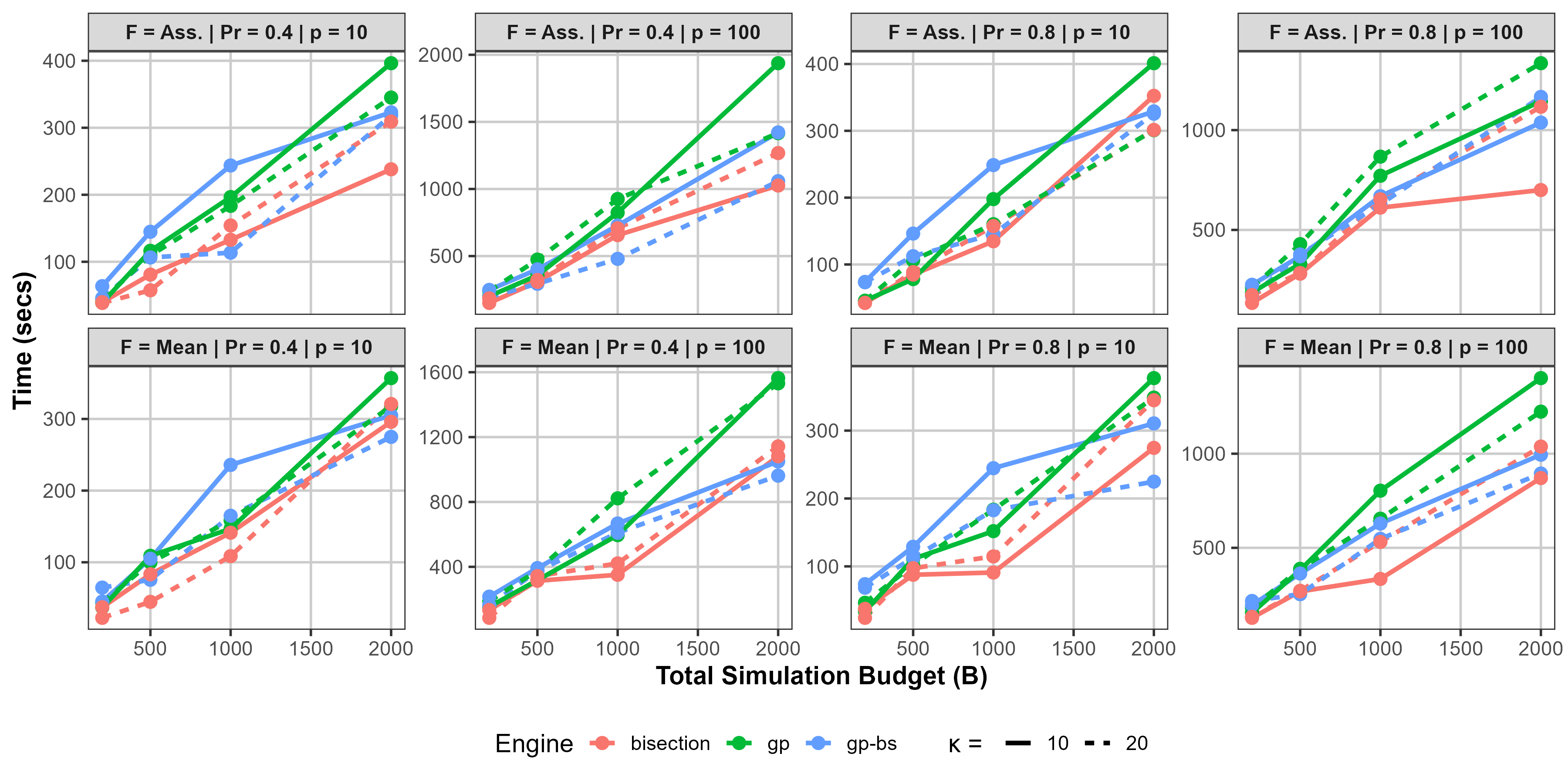}
        \caption{Computational time required to estimate minimum sample size $n^*$.}
        \label{fig:aim1-surv-time}
    \end{subfigure}
    \caption{Aim~1 (Survival outcome): Comparison of CV and computational time across search engines \texttt{gp}, \texttt{bisection} and \texttt{gp-bs} for calibration slope metric under varying $(\kappa = 10, 20)$, event rate $(0.4, 0.8)$, number of predictors $(p = 10, 100)$ and total simulation budget $(B = 200$--$2000)$.}
    \label{fig:aim1-survival-summary}
\end{figure}

Figure~\ref{fig:aim1-survival-summary} compares sample size precision (CV) and computational time for the calibration slope metric across design configurations: $\kappa \in \{10, 20\}$, event rate ($0.4, 0.8$), $p \in \{10, 100\}$, and $B = 200$--$2000$. The \texttt{gp} engine consistently yielded lower and more stable CV values than both \texttt{bisection} and \texttt{gp-bs}, with the relative advantage most marked in the more challenging low-event-rate scenarios ($0.4$). Performance improvements were more systematic under mean-based aggregation. CV values approached a plateau at $B \approx 1000$ for $\kappa = 20$ under both aggregation schemes. The corresponding results for the C-index metric are reported in Figure~\ref{fig:aim1-survival-summary2} (Additional~File~1).
 
Under $B = 1000$ (Table~\ref{tab:survival_combined}), \texttt{gp} consistently achieved the lowest CV for C-index estimation, often by a substantial margin, with the advantage most evident in low-event-rate ($0.4$), high-dimensional ($p = 100$) scenarios. The only exception arose under assurance-based aggregation with $p = 10$ and event rate $0.4$, where \texttt{gp-bs} produced a slightly smaller CV. Calibration slope estimation was more demanding, yet \texttt{gp} still consistently achieved the lowest CV across all scenarios. The \texttt{bisection} procedure systematically underperformed under nearly all evaluated conditions.
 
\begin{table}[H]
\centering
\caption{\label{tab:survival_combined}
Estimated minimum sample size $\hat{n}^*$ and stability (CV) for survival outcomes using assurance aggregation ($20\%$ quantile) and mean aggregation, stratified by target metric, event rate and number of predictors ($p$), at $B = 1000$ and $\kappa = 20$. Results are averaged over 100 simulation replicates.}
\centering
\fontsize{8}{9}\selectfont
\begin{tabular}[t]{llllcccc}
\toprule
\textbf{Metric} & \textbf{Event rate} & $\bm p$ & \textbf{Engine} & \multicolumn{2}{c}{\textbf{Mean} $\hat{n}^*$} & \multicolumn{2}{c}{\textbf{CV}} \\
 &  &  &  & Assurance & Mean agg. & Assurance & Mean agg. \\
\midrule
\multirow{12}{*}[1.5\dimexpr\aboverulesep+\belowrulesep+\cmidrulewidth]{\raggedright\arraybackslash C-index}
 &  &  & bisection & 60 & 47 & 11.74 & 10.80 \\
 &  &  & gp & 62 & \textbf{48} & 8.48 & \textbf{4.77} \\
 &  & \multirow{-3}{*}{\raggedright\arraybackslash 10} & \textbf{gp-bs} & \textbf{62} & 48 & \textbf{6.77} & 4.82 \\
\cmidrule{3-8}
 &  &  & bisection & 540 & 489 & 4.55 & 3.58 \\
 &  &  & \textbf{gp} & \textbf{543} & \textbf{491} & \textbf{2.38} & \textbf{0.47} \\
 & \multirow{-6}{*}[0.5\dimexpr\aboverulesep+\belowrulesep+\cmidrulewidth]{\raggedright\arraybackslash 0.4}
   & \multirow{-3}{*}{\raggedright\arraybackslash 100} & gp-bs & 541 & 491 & 3.22 & 2.44 \\
\cmidrule{2-8}
 &  &  & bisection & 38 & 31 & 12.05 & 10.71 \\
 &  &  & \textbf{gp} & \textbf{41} & \textbf{32} & \textbf{4.69} & \textbf{4.43} \\
 &  & \multirow{-3}{*}{\raggedright\arraybackslash 10} & gp-bs & 40 & 32 & 4.78 & 3.65 \\
\cmidrule{3-8}
 &  &  & bisection & 334 & 304 & 5.04 & 3.12 \\
 &  &  & \textbf{gp} & \textbf{334} & \textbf{304} & \textbf{0.86} & \textbf{0.48} \\
 & \multirow{-6}{*}[0.5\dimexpr\aboverulesep+\belowrulesep+\cmidrulewidth]{\raggedright\arraybackslash 0.8}
   & \multirow{-3}{*}{\raggedright\arraybackslash 100} & gp-bs & 334 & 304 & 2.39 & 1.97 \\
\midrule[\heavyrulewidth]
\multirow{12}{*}[1.5\dimexpr\aboverulesep+\belowrulesep+\cmidrulewidth]{\raggedright\arraybackslash Calib.\ Slope}
 &  &  & bisection & 486 & 268 & 25.79 & 20.96 \\
 &  &  & \textbf{gp} & \textbf{479} & \textbf{270} & \textbf{21.00} & \textbf{12.94} \\
 &  & \multirow{-3}{*}{\raggedright\arraybackslash 10} & gp-bs & 471 & 267 & 17.49 & 12.67 \\
\cmidrule{3-8}
 &  &  & bisection & 3524 & 2992 & 11.84 & 8.00 \\
 &  &  & \textbf{gp} & \textbf{3403} & \textbf{2990} & \textbf{5.92} & \textbf{3.78} \\
 & \multirow{-6}{*}[0.5\dimexpr\aboverulesep+\belowrulesep+\cmidrulewidth]{\raggedright\arraybackslash 0.4}
   & \multirow{-3}{*}{\raggedright\arraybackslash 100} & gp-bs & 3504 & 2941 & 8.37 & 6.53 \\
\cmidrule{2-8}
 &  &  & bisection & 328 & 178 & 22.46 & 17.43 \\
 &  &  & \textbf{gp} & \textbf{324} & \textbf{171} & \textbf{10.53} & \textbf{5.63} \\
 &  & \multirow{-3}{*}{\raggedright\arraybackslash 10} & gp-bs & 315 & 175 & 12.86 & 11.48 \\
\cmidrule{3-8}
 &  &  & bisection & 2277 & 1863 & 9.07 & 7.85 \\
 &  &  & \textbf{gp} & \textbf{2198} & \textbf{1846} & \textbf{5.36} & \textbf{3.88} \\
 & \multirow{-6}{*}[0.5\dimexpr\aboverulesep+\belowrulesep+\cmidrulewidth]{\raggedright\arraybackslash 0.8}
   & \multirow{-3}{*}{\raggedright\arraybackslash 100} & gp-bs & 2225 & 1878 & 6.78 & 5.06 \\
\bottomrule
\end{tabular}
\end{table}

\subsubsection*{Overall ranking}
 
Table~\ref{tab:overall_ranking} summarises the comparative performance of the three search algorithms across all 12 outcome--aggregation--metric configurations. \texttt{gp} achieves the lowest mean rank in 11 of the 12 configurations and is overall the best-performing search engine, achieving a perfect average rank of 1.00 in nine configurations. The sole exception arises for the calibration slope under binary outcome with assurance-based aggregation, where \texttt{gp-bs} attains the best rank. The \texttt{bisection} procedure is ranked last or tied for last in every configuration, consistent with the inefficiency of deterministic bisection in stochastic optimisation settings.
 
\begin{table}[H]
\centering
\caption{Average rank of search engines across outcomes, aggregation methods, and target metrics at $B = 1000$. Lower average rank indicates better overall performance (CV).}
\label{tab:overall_ranking}
\begin{tabular}{lllccc}
\toprule
Outcome    & Aggregation method & Metric        & bisection & gp & gp-bs \\
\midrule
     & Assurance          & AUC           & 3.00      & \textbf{1.00} & 2.00 \\
Binary     &           & Calib. Slope  & 2.75      & 1.75  & \textbf{1.50} \\
\cmidrule{2-6}
     & Mean               & AUC           & 3.00      & \textbf{1.00} & 2.00 \\
     &                & Calib. Slope  & 2.75      & \textbf{1.25} & 2.00 \\
\midrule[\heavyrulewidth]
 & Assurance          & $R^2$         & 2.75      & \textbf{1.00} & 2.25 \\
Continuous &          & Calib. Slope  & 3.00      & \textbf{1.00} & 2.00 \\
\cmidrule{2-6}
 & Mean               & $R^2$         & 2.75      & \textbf{1.00} & 2.25 \\
 &               & Calib. Slope  & 3.00      & \textbf{1.00} & 2.00 \\
 \midrule[\heavyrulewidth]
   & Assurance          & C-index       & 2.50      & \textbf{1.25} & 2.25 \\
Survival   &           & Calib. Slope  & 2.75      & \textbf{1.00} & 2.25 \\
\cmidrule{2-6}
   & Mean               & C-index       & 3.00      & \textbf{1.00} & 2.00 \\
   &               & Calib. Slope  & 2.50      & \textbf{1.00} & 2.50 \\
\bottomrule
\end{tabular}
\end{table}

\subsection{Aim~2: Benchmark comparison}
 
\subsubsection*{Binary outcome}
 
Table~\ref{tab:binary_perf_dev} reports the percentage deviation of achieved performance from target for binary outcome models with $p = 20$. \texttt{pmsims} (mean) and \texttt{samplesizedev} exhibited the most stable performance, with deviations typically within $\pm 1\%$ of the target. \texttt{pmsampsize} generally suggested the smallest sample sizes; although it occasionally achieved performance close to the target, it showed substantial negative deviations in several scenarios, most pronounced at the highest target discrimination (AUC $0.9$), where deviations reached $-7.24\%$ (prevalence 5\%) and $-9.84\%$ (prevalence 20\%). \texttt{pmsims} (assurance) recommended the largest sample sizes and produced consistently minor negative deviations (ranging from $-1.05\%$ to $-2.00\%$). Supplementary figures showing minimum sample size requirements, coefficient of variation, computational time, and achieved calibration performance across all predictor counts and scenarios are provided in Figures~\ref{fig:aim2-min-n-bin}--\ref{fig:aim2-perf-bin} (Additional~File~1).
 
\begin{table}[H]
\caption{\label{tab:binary_perf_dev}Comparison of sample size requirements and achieved performance deviations for binary outcomes across two prevalence levels (0.05, 0.20) and two target large-sample AUC values (0.8, 0.9), with $p = 20$. Performance evaluated for calibration slope.}
\fontsize{8}{9}\selectfont
\centering
\begin{tabular}[t]{llclccc}
\toprule
Metric & Prevalence & Large-sample $AUC$ & Engine & $\hat{n}^*$ & $\hat{G}(\hat{n}^*)$ & \% Deviation \\
\midrule[\heavyrulewidth]
&  &  & \texttt{pmsims} (assurance) & 5562 & 0.882\,  & -2.00\, \\
& 0.05 & 0.8 & \texttt{pmsims} (mean) & 3403 & 0.906\,  & 0.70\, \\
&  &  & \texttt{pmsampsize} & 2788 & 0.888\,  & -1.34\, \\
&  &  & \texttt{samplesizedev}  & 3346 & 0.903\,  & 0.34\, \\
\cmidrule{3-7}
&  &  & \texttt{pmsims} (assurance) & 4080 & 0.890\,  & -1.16\, \\
&  & 0.9 & \texttt{pmsims} (mean) & 2352 & 0.911\,  & 1.18\, \\
&  &  & \texttt{pmsampsize} & 1303 & 0.835\,  & -7.24\, \\
&  &  & \texttt{samplesizedev}  & 2331 & 0.908\,  & 0.91\, \\
\cmidrule{2-7}
Calib. slope&  &  & \texttt{pmsims} (assurance) & 2024 & 0.884\,  & -1.75\, \\
&  0.20 & 0.8 & \texttt{pmsims} (mean) & 1239 & 0.898\,  & -0.25\, \\
&  &  & \texttt{pmsampsize}  & 882 & 0.863\,  & -4.14\, \\
&  &  & \texttt{samplesizedev}   & 1245 & 0.902\,  & 0.21\, \\
\cmidrule{3-7}
&  &  & \texttt{pmsims} (assurance) & 1674 & 0.891\,  & -1.05\, \\
&  & 0.9 & \texttt{pmsims} (mean) & 986 & 0.898\,  & -0.24\, \\
&  &  & \texttt{pmsampsize}   & 509 & 0.811\,  & -9.84\, \\
&  &  & \texttt{samplesizedev}   & 1005 & 0.904\,  & 0.42\, \\
\bottomrule
\end{tabular}
\end{table}
 
\subsubsection*{Continuous outcome}
 
Table~\ref{tab:continuous_perf_dev} presents the percentage deviation of observed performance from prespecified targets for continuous outcome models with $p = 20$. \texttt{pmsims} (mean) achieved high precision, with deviations within $\pm 0.09\%$ at $R^2 = 0.5$ and $0.7$, and a modest positive deviation of $+0.98\%$ at $R^2 = 0.2$. The assurance-based variant produced slight negative deviations ($-0.12\%$ to $-0.42\%$). In contrast, \texttt{pmsampsize} showed systematic positive deviations that increased strongly with target $R^2$: from $+0.93\%$ at $R^2 = 0.2$ to $+7.53\%$ at $R^2 = 0.7$, reflecting the fact that in higher-signal settings the \texttt{pmsampsize} sample size is not driven by shrinkage, resulting in calibration slopes that exceed the target. Supplementary figures are provided in Figures~\ref{fig:aim2-min-n-cont}--\ref{fig:aim2-perf-cont} (Additional~File~1).
 
\begin{table}[H]
\caption{\label{tab:continuous_perf_dev}Comparison of sample size requirements and achieved performance deviations for continuous outcomes across three target $R^2$ levels (0.2, 0.5, 0.7), with $p = 20$. Performance evaluated for calibration slope.}
\fontsize{8}{9}\selectfont
\centering
\begin{tabular}[t]{lclccc}
\toprule
Metric & Large-sample $R^2$ & Engine & $\hat{n}^*$ & $\hat{G}(\hat{n}^*)$ & \% Deviation\\
\midrule[\heavyrulewidth]
&  & \texttt{pmsims} (assurance) & 1196 & 0.897\,  & -0.39\,  \\
& 0.2 & \texttt{pmsims} (mean) & 702 & 0.909\,  & 0.98\,  \\
&  & \texttt{pmsampsize}  & 716 & 0.908\,  & 0.93\,  \\
\cmidrule{2-6}
&  & \texttt{pmsims} (assurance) & 320 & 0.899\,  & -0.12\,  \\
Calib. slope & 0.5 & \texttt{pmsims} (mean) & 189 & 0.899\,  & -0.09\,  \\
&  & \texttt{pmsampsize} & 254 & 0.928\,  & 3.09\,  \\
\cmidrule{2-6}
&  & \texttt{pmsims} (assurance) & 151 & 0.896\,  & -0.42\,  \\
& 0.7 & \texttt{pmsims} (mean) & 95 & 0.900\,  & 0.03\,  \\
&  & \texttt{pmsampsize} & 254 & 0.968\,  & 7.53\,  \\
\bottomrule
\end{tabular}
\end{table}
 
\subsubsection*{Survival outcome}
 
Table~\ref{tab:survival_perf_dev} compares the sample size requirements and achieved calibration slope deviations for survival outcome models with $p = 20$. All three methods achieved calibration slopes close to the target of $0.90$ across the full range of event rates and target C-index values. \texttt{pmsims} (mean) produced the smallest deviations overall, consistently within $\pm 0.87\%$ across all scenarios. \texttt{pmsims} (assurance) showed modest negative deviations, ranging from $-0.41\%$ to $-1.47\%$, reflecting its conservative sample size recommendations. \texttt{samplesizedev} performed comparably at event rates of $0.4$ and $0.5$, with deviations between $-0.94\%$ and $+3.19\%$, though positive deviations at $C$-index $= 0.9$ (up to $+3.19\%$ at event rate $0.4$ and $+2.78\%$ at event rate $0.5$) suggest mild overestimation of the required sample size in high-discrimination settings. At event rate $0.8$, all three methods performed well, with deviations within $\pm 1.67\%$. Supplementary figures are provided in Figures~\ref{fig:aim2-min-n-surv}--\ref{fig:aim2-perf-surv} (Additional~File~1).
 
\begin{table}[H]
\caption{\label{tab:survival_perf_dev}Comparison of sample size requirements and achieved performance deviations for survival outcomes across three event rates (0.4, 0.5, 0.8) and three target $C$-index values (0.7, 0.8, 0.9), with $p = 20$. Performance evaluated for calibration slope.}
\centering
\fontsize{8}{9}\selectfont
\begin{tabular}[t]{lcclccc}
\toprule
Metric & Event rate & Large-sample $C$-index & Engine & $\hat{n}^*$ & $\hat{G}(\hat{n}^*)$ & \% Deviation \\
\midrule[\heavyrulewidth]
&  &  & \texttt{pmsims} (assurance) & 1569 & 0.888\,  & -1.34\, \\
&  & 0.7 & \texttt{pmsims} (mean) & 963 & 0.896\,  & -0.41\, \\
&  &  & \texttt{samplesizedev}  & 925 & 0.892\,  & -0.94\, \\
\cmidrule{3-7}
&  &  & \texttt{pmsims} (assurance) & 889 & 0.891\,  & -1.03\, \\
& 0.4 & 0.8 & \texttt{pmsims} (mean) & 584 & 0.908\,  & 0.87\, \\
&  &  & \texttt{samplesizedev}  & 544 & 0.895\,  & -0.52\, \\
\cmidrule{3-7}
&  &  & \texttt{pmsims} (assurance) & 680 & 0.893\,  & -0.79\, \\
&  & 0.9 & \texttt{pmsims} (mean) & 478 & 0.901\,  & 0.16\, \\
&  &  & \texttt{samplesizedev}  & 656 & 0.929\,  & 3.19\, \\
\cmidrule{2-7}
&  &  & \texttt{pmsims} (assurance) & 1283 & 0.887\,  & -1.47\, \\
&  & 0.7 & \texttt{pmsims} (mean) & 808 & 0.902\,  & 0.26\, \\
&  &  & \texttt{samplesizedev}  & 802 & 0.899\,  & -0.11\, \\
\cmidrule{3-7}
&  &  & \texttt{pmsims} (assurance) & 756 & 0.894\,  & -0.72\, \\
Calib. slope & 0.5 & 0.8 & \texttt{pmsims} (mean) & 493 & 0.898\,  & -0.21\, \\
&  &  & \texttt{samplesizedev}  & 472 & 0.897\,  & -0.33\, \\
\cmidrule{3-7}
&  &  & \texttt{pmsims} (assurance) & 590 & 0.896\,  & -0.41\, \\
&  & 0.9 & \texttt{pmsims} (mean) & 408 & 0.903\,  & 0.33\, \\
&  &  & \texttt{samplesizedev}  & 535 & 0.925\,  & 2.78\, \\
\cmidrule{2-7}
&  &  & \texttt{pmsims} (assurance) & 919 & 0.891\,  & -1.01\, \\
&  & 0.7 & \texttt{pmsims} (mean) & 607 & 0.901\,  & 0.10\, \\
&  &  & \texttt{samplesizedev}  & 604 & 0.899\,  & -0.11\, \\
\cmidrule{3-7}
&  &  & \texttt{pmsims} (assurance) & 574 & 0.901\,  & 0.12\, \\
& 0.8 & 0.8 & \texttt{pmsims} (mean) & 353 & 0.895\,  & -0.60\, \\
&  &  & \texttt{samplesizedev}  & 374 & 0.901\,  & 0.11\, \\
\cmidrule{3-7}
&  &  & \texttt{pmsims} (assurance) & 436 & 0.896\,  & -0.48\, \\
&  & 0.9 & \texttt{pmsims} (mean) & 292 & 0.901\,  & 0.15\, \\
&  &  & \texttt{samplesizedev}  & 340 & 0.915\,  & 1.67\, \\
\bottomrule
\end{tabular}
\end{table}

\section{Discussion}

Researchers developing clinical prediction models face the critical challenge of justifying their chosen sample sizes, yet practical guidance on which sample size calculation method to use remains limited. In this study, we evaluated three simulation-based search algorithms implemented in the \texttt{pmsims} package---\texttt{gp} (Gaussian process-based), a deterministic bisection procedure, and a hybrid \texttt{gp-bs} approach---for estimating the minimum development sample size required for prediction models with binary, continuous, and time-to-event outcomes. We assessed each method under both mean-based and assurance-based criteria across a wide range of scenarios varying outcome prevalence or event rate, predictor dimensionality, and target performance metrics. We also benchmarked the best-performing method against established R packages for sample size calculation.

\subsection{Simulation findings and recommendations}

In all scenarios, the GP-based \texttt{gp} search engine produced the most precise estimates of the minimum sample size (lowest CV) when compared with a classical bisection search and the hybrid \texttt{gp-bs} approach. The advantage was especially large in challenging scenarios characterised by low signal (low $R^2$ or low event prevalence) and high dimensionality. This finding aligns with recent methodological advances in stochastic root-finding, where adaptive sampling methods like the Probabilistic Bisection Algorithm maintain near-optimal convergence rates even in low-signal regimes \cite{chalmers2024solving,yu2025root}.

Increasing the number of simulation replicates per candidate sample size ($\kappa$) substantially stabilised the surrogate model updates and reduced the CV of the estimated $n^*$; $\kappa = 20$ (rather than $\kappa = 10$) yielded markedly lower CV for \texttt{gp}. A total simulation budget of roughly $B \approx 1000$ was an efficient trade-off between precision and cost, with only modest performance improvements beyond this threshold. This efficiency gain reflects the sequential nature of the GP-based search, which automatically concentrates simulation effort in regions where greater precision is needed \cite{yu2025root}.

As expected, the assurance criterion produced more conservative recommended sample sizes than mean-based targets. \texttt{gp} produced reliable assurance-based recommendations but required higher simulation effort to achieve comparable precision to mean-based estimates. The hybrid \texttt{gp-bs} sometimes achieved marginally better precision than \texttt{gp} for calibration targets under assurance in limited binary settings, reflecting the benefit of a coarse deterministic range-finding step when the performance curve is particularly noisy.

The deterministic bisection search consistently underperformed, requiring many more model evaluations to achieve comparable stability. This supports the view that deterministic interval halving is not well suited to stochastic sample size searches unless a very large evaluation budget is available \cite{yu2025root}.

When compared with \texttt{pmsampsize} (analytical) and \texttt{samplesizedev} (simulation-based), the \texttt{pmsims} \texttt{gp} engine produced recommended sample sizes that achieved the prespecified discrimination and calibration targets when evaluated on large independent validation sets. Analytical formulae can be sensitive to modelling assumptions and may under- or overestimate the required sample sizes in some high model strength settings \cite{Riley2020}, whereas simulation-based approaches can more directly reflect performance variability under the assumed data-generating mechanism. Our findings suggest that the GP-based adaptive search implemented in \texttt{gp} offers a compelling middle ground: it achieves the precision of simulation-based approaches while maintaining comparable sample efficiency through sequential learning.

\subsection{Limitations and implications for future research}

Our validation study focused on data-generating mechanisms with continuous predictors and seven levels of model dimensionality ($p = 5, \dots, 100$), where all candidate predictors corresponded to true signal variables. This design does not fully capture real-world complexities such as mixtures of discrete and continuous covariates, correlated noise features, missing data, or heavy-tailed distributions. Additional data-generating mechanisms incorporating realistic correlation structures and explicit noise predictors will be required to more fully characterise the generalisability of the findings.

The specification of the starting value can influence numerical stability. Analytical approaches such as \texttt{pmsampsize} use closed-form formulae and are insensitive to starting values. By contrast, simulation-based procedures can be sensitive to their initial search region. The \texttt{pmsims} \texttt{gp} engine uses adaptive, data-driven starting values (see \autoref{startvalues}) to focus the search on informative regions of the sample size space; early iterations may be unstable when the performance curve is flat or multimodal.

\section{Conclusion}

The \texttt{pmsims} package uses a GP-based \texttt{gp} search engine to provide a computationally efficient and flexible framework for principled sample size planning in clinical prediction modelling. Our evaluation shows that adaptive surrogate-based search substantially outperforms deterministic bisection approaches and achieves comparable performance relative to established analytical and simulation-based tools while requiring fewer evaluations. With pragmatic adaptive starting limits, sufficient replication per evaluation ($\kappa \ge 20$), and transparent reporting of uncertainty around the estimated minimum $n$, simulation-based planning can improve the reliability of prediction models developed in practice.

Future work should prioritise: (1) more robust surrogate modelling under non-monotonic or highly variable performance curves, (2) improved assurance-based estimation to reduce simulation budgets for quantile targets, and (3) broader validation across more realistic data-generating mechanisms and model classes, including correlated covariates, noise predictors, missing data, and predictive uncertainty.

\backmatter

\bmhead{Supplementary information}

Additional~File~1 contains supplementary figures for the Aim~1 and Aim~2 results, including CV and computational time plots for the AUC metric (binary), $R^2$ and calibration slope metrics (continuous), and C-index metric (survival) from the search engine comparisons, as well as minimum sample size requirements, coefficient of variation, computational time, and achieved calibration performance figures from the benchmark comparisons with \texttt{pmsampsize} and \texttt{samplesizedev} for binary, continuous, and survival outcomes.

\bmhead{Acknowledgements}

We are grateful to the Advisory Group of the \texttt{pmsims} project for their sustained guidance and support throughout the development of this work. We acknowledge the contributions of Dr Nick Cummins (King’s College London, UK) and Dr Joie Ensor (University of Birmingham, UK), whose academic insights strengthened both the conceptual and methodological foundations of the project. We also extend our thanks to public representatives Katherine Barrett, Emma Shellard, and Aurora Todisco, whose lived experience and thoughtful engagement helped ensure that the development of the package remained grounded in real-world needs and user priorities. Their combined expertise significantly enriched the direction and clarity of this work.

\section*{Declarations}

\textbf{Funding.} 
The \texttt{pmsims} project, including this study, is funded by the NIHR Research for Patient Benefit programme (NIHR206858). The views expressed are those of the authors and not necessarily those of the NIHR or the Department of Health and Social Care. ORO is supported by NIHR206858 at King’s College London. This study is part-funded by the National Institute for Health and Care Research (NIHR) Maudsley Biomedical Research Centre (BRC). DSt, DSh, GF, and EC received financial support from the National Institute for Health Research (NIHR) Biomedical Research Centre at South London and Maudsley NHS Foundation Trust (NIHR203318) and King's College London.

\noindent \textbf{Competing interests.} The authors declare that they have no competing interests.

\noindent \textbf{Ethics approval and consent to participate.} Not applicable. This study involved no human participants, human data, or animal subjects; all data were generated by computer simulation.

\noindent \textbf{Consent for publication.} Not applicable.

\noindent \textbf{Data availability.} No empirical datasets were generated or analysed in this study. All simulation results are reproducible using the \texttt{pmsims} R package (version~0.5.0) with the simulation scenarios and parameter settings described in the Methods section.

\noindent \textbf{Materials availability.} Not applicable.

\noindent \textbf{Code availability.} The \texttt{pmsims} R package (version~0.5.0) is freely available on GitHub at \url{https://github.com/pmsims-package/pmsims/}. All simulation code required to reproduce the results presented in this paper is available from the corresponding author upon request.

\noindent \textbf{Author contributions.} ORO, GF, and EC designed the simulation study. ORO conducted the simulation study, analysed the results, and drafted the manuscript. ORO, DSh, FZ, GF, and EC developed the \texttt{pmsims} package. DSh, SM, FZ, DSt, GF, and EC contributed to the conceptualisation and study design, interpretation of results, and critical revision of the manuscript. GF and EC provided oversight of the project. All authors read and approved the final manuscript.

\begin{appendices}
 
\renewcommand{\thefigure}{S\arabic{figure}}
\setcounter{figure}{0}
 
\section{Additional File 1: Supplementary Figures}\label{secA1}
 
\subsection*{Aim~1 Supplementary Figures}
 
\begin{figure}[H]
    \centering
    \begin{subfigure}[t]{0.95\linewidth}
        \centering
        \includegraphics[width=\linewidth]{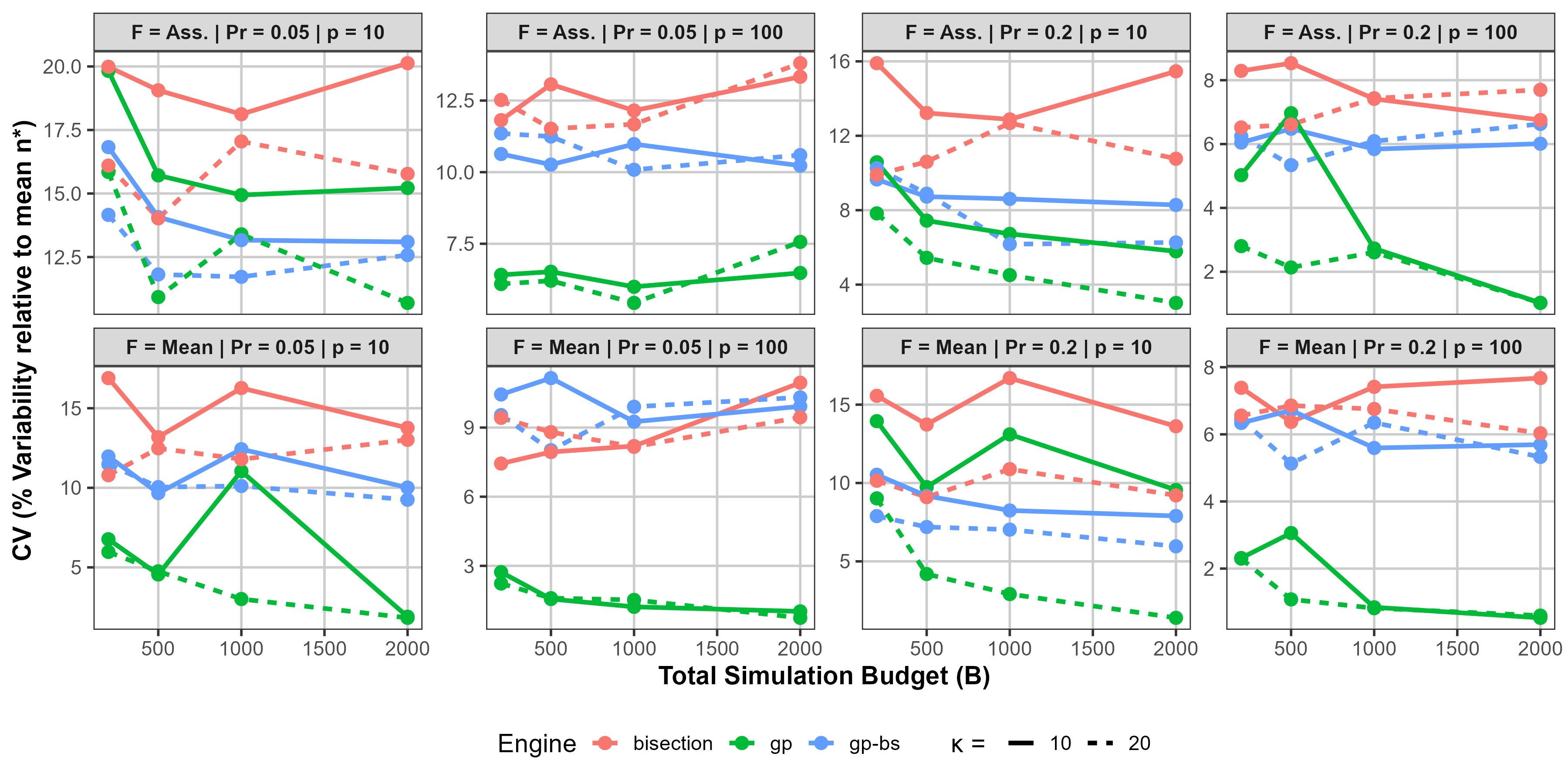}
        \caption{Coefficient of Variation (CV) of sample size estimates relative to mean $n^*$.}
        \label{fig:aim1-cv2}
    \end{subfigure}
    \hspace{0.5cm}
    \begin{subfigure}[t]{0.95\linewidth}
        \centering
        \includegraphics[width=\linewidth]{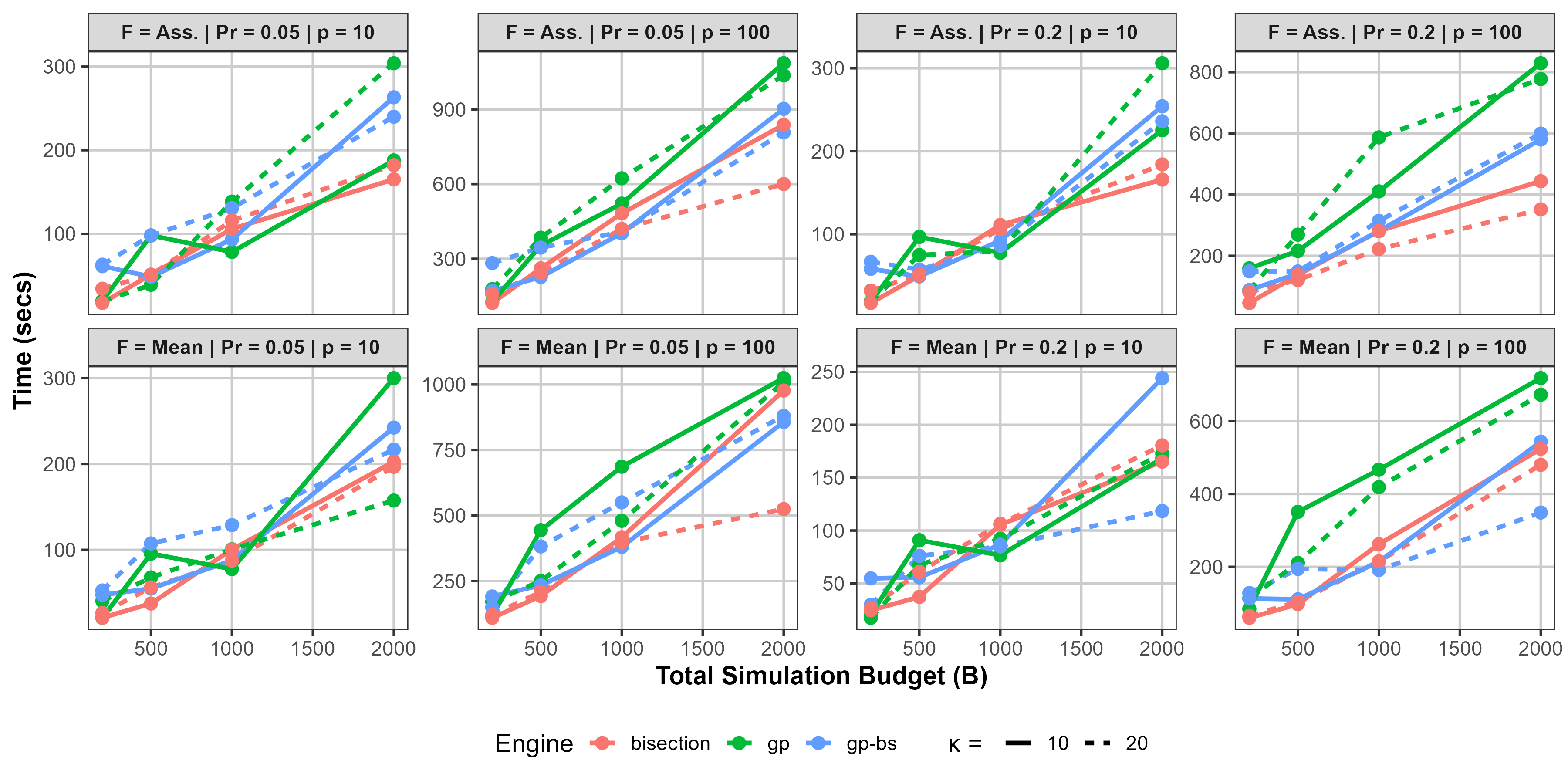}
        \caption{Computational time required to estimate minimum sample size $n^*$.}
        \label{fig:aim1-time2}
    \end{subfigure}
    \caption{Aim~1 (Binary outcome): Comparison of CV and computational time across search engines \texttt{gp}, \texttt{bisection} and \texttt{gp-bs} for AUC metric under varying $(\kappa = 10, 20)$, prevalence $(0.05, 0.20)$, number of predictors $(p = 10, 100)$ and total simulation budget $(B = 200$--$2000)$.}
    \label{fig:aim1-binary-summary2}
\end{figure}
 
\begin{figure}[H]
    \centering
    \begin{subfigure}[t]{0.95\linewidth}
        \centering
        \includegraphics[width=\linewidth]{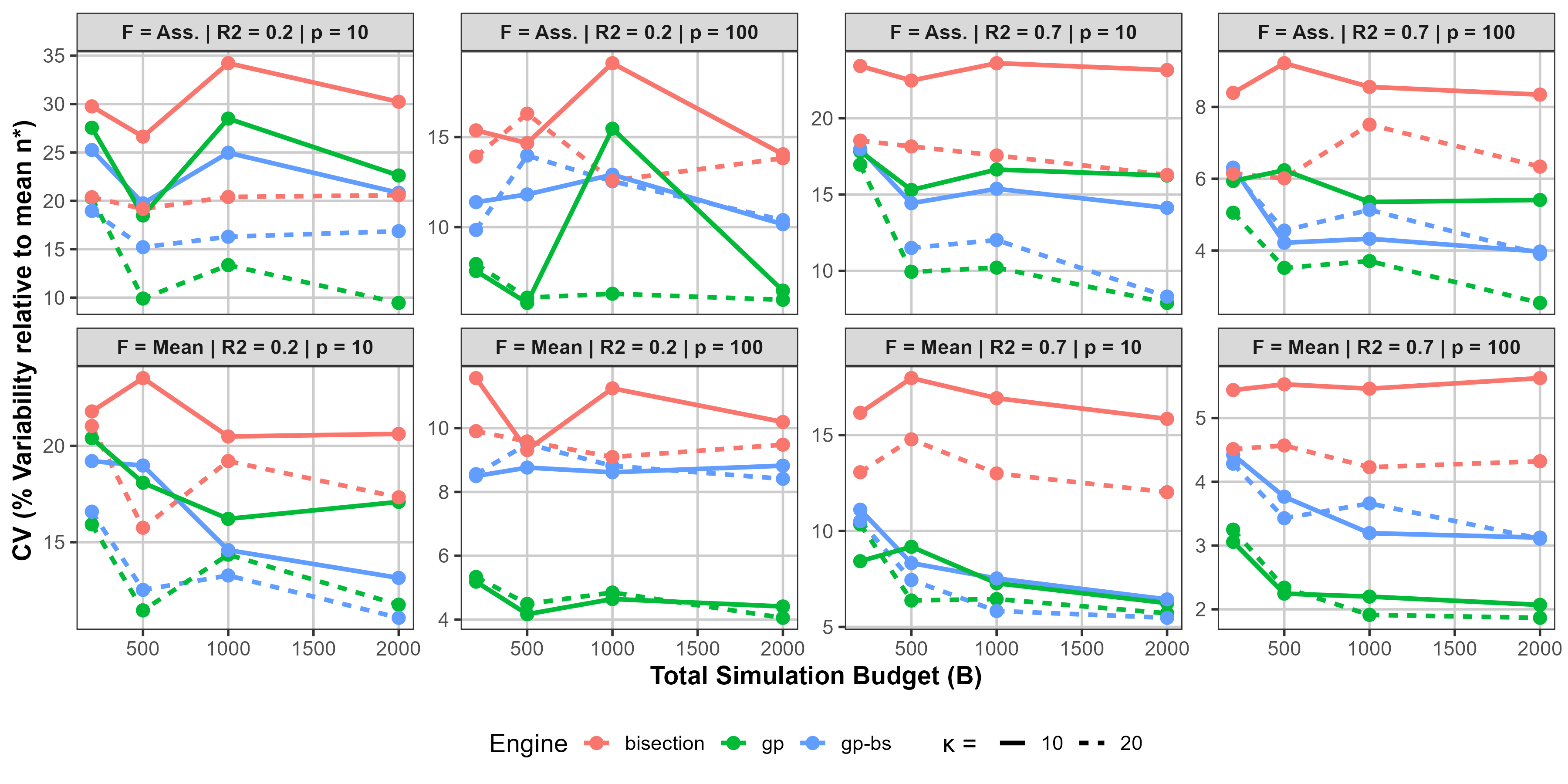}
        \caption{Coefficient of Variation (CV) of sample size estimates relative to mean $n^*$.}
        \label{fig:aim1-cont-cv}
    \end{subfigure}
    \hspace{0.5cm}
    \begin{subfigure}[t]{0.95\linewidth}
        \centering
        \includegraphics[width=\linewidth]{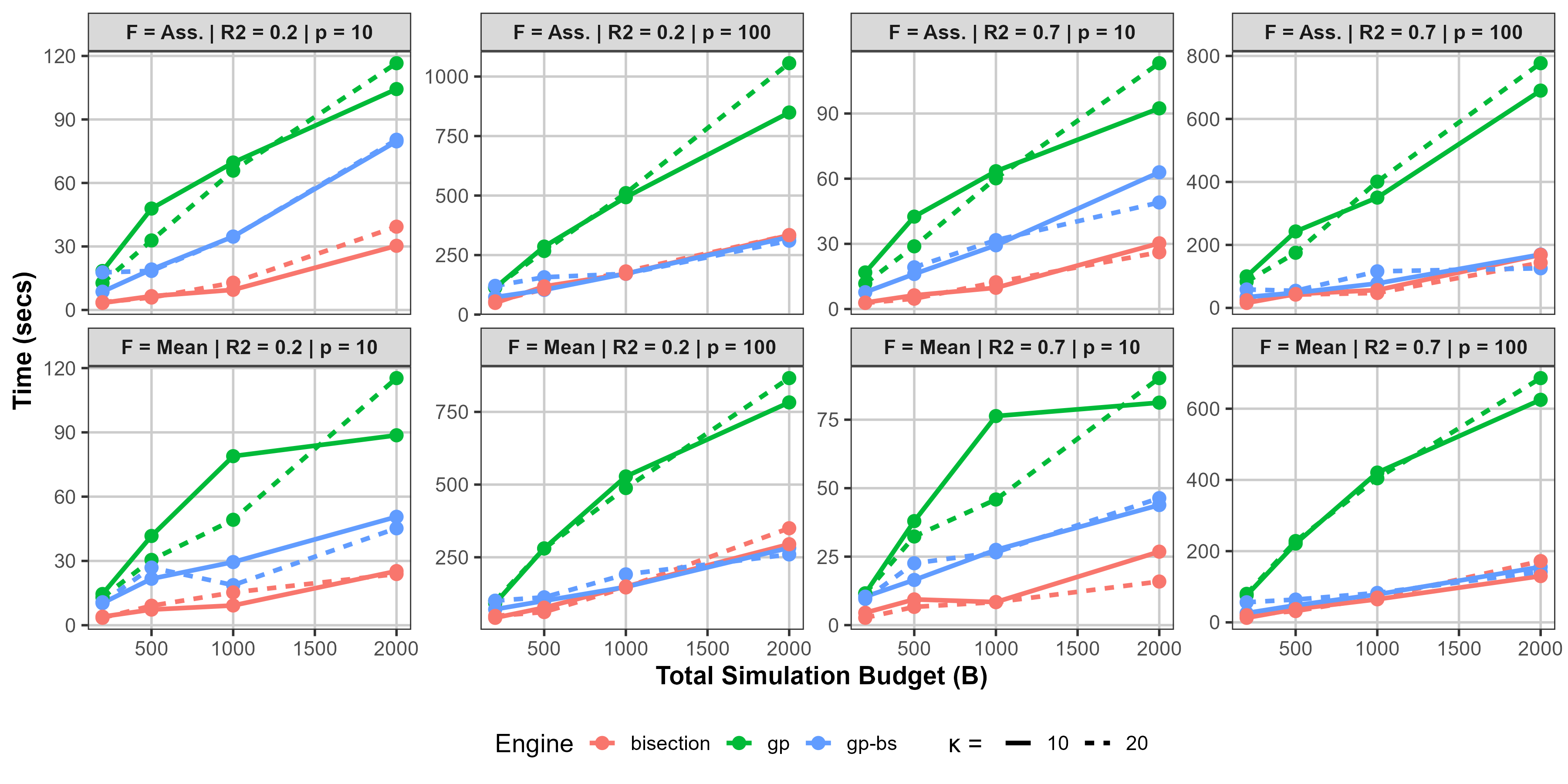}
        \caption{Computational time required to estimate minimum sample size $n^*$.}
        \label{fig:aim1-cont-time}
    \end{subfigure}
    \caption{Aim~1 (Continuous outcome): Comparison of CV and computational time across search engines \texttt{gp}, \texttt{bisection} and \texttt{gp-bs} for calibration slope metric under varying $(\kappa = 10, 20)$, $R^2 = 0.2, 0.7$, number of predictors $(p = 10, 100)$ and total simulation budget $(B = 200$--$2000)$.}
    \label{fig:aim1-continuous-summary}
\end{figure}
 
\begin{figure}[H]
    \centering
    \begin{subfigure}[t]{0.95\linewidth}
        \centering
        \includegraphics[width=\linewidth]{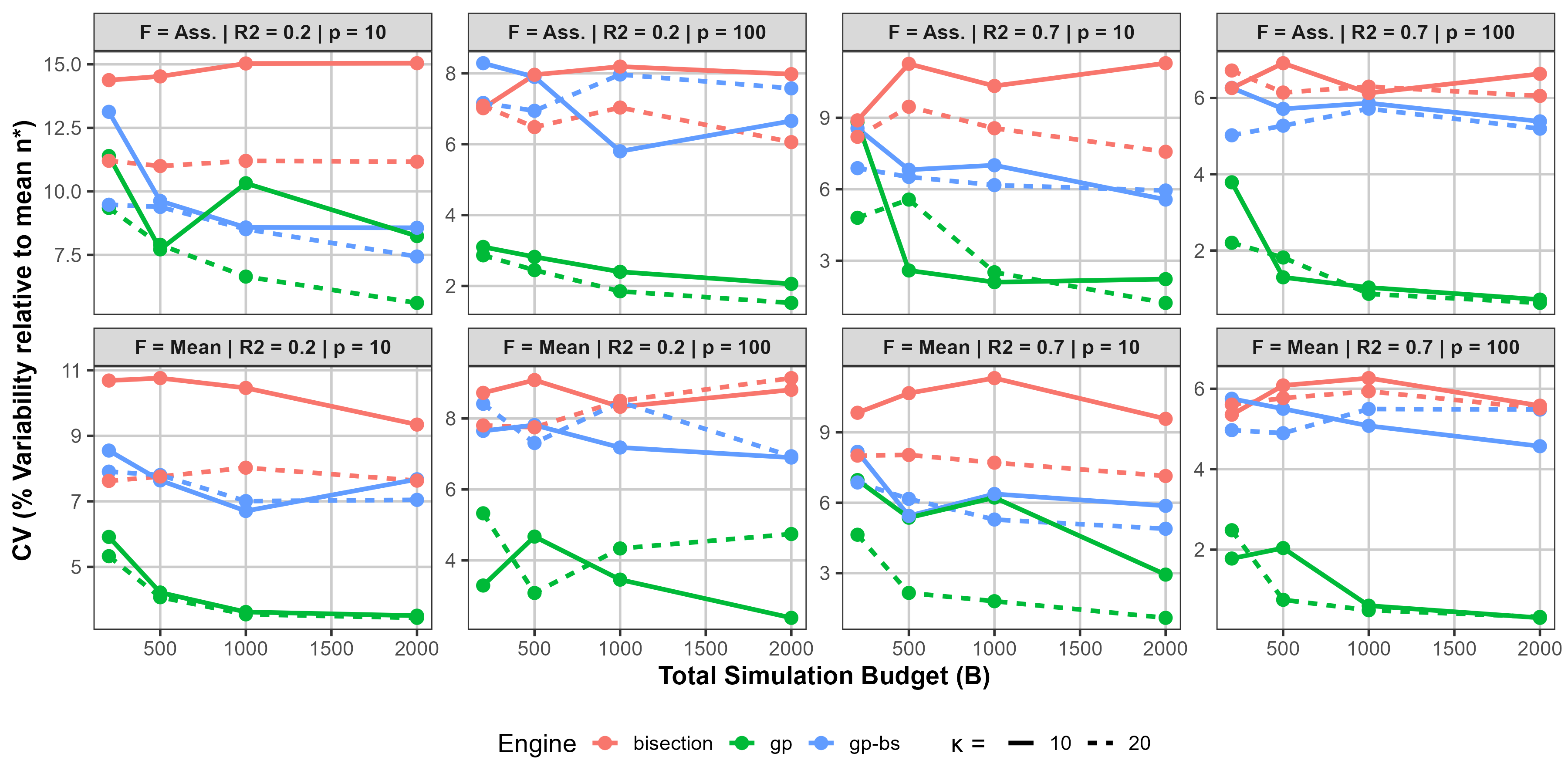}
        \caption{Coefficient of Variation (CV) of sample size estimates relative to mean $n^*$.}
        \label{fig:aim1-cont-cv2}
    \end{subfigure}
    \hspace{0.5cm}
    \begin{subfigure}[t]{0.95\linewidth}
        \centering
        \includegraphics[width=\linewidth]{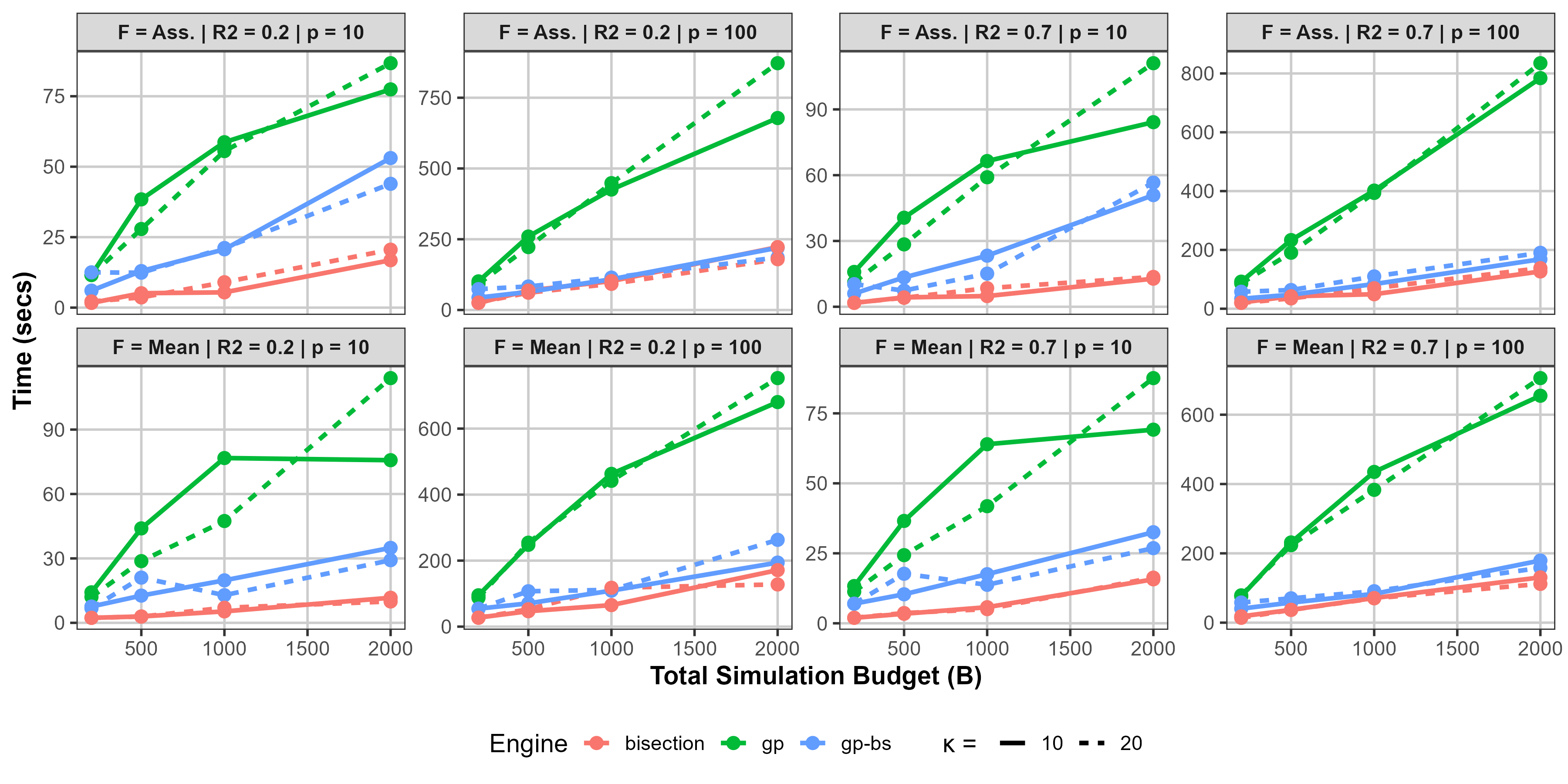}
        \caption{Computational time required to estimate minimum sample size $n^*$.}
        \label{fig:aim1-cont-time2}
    \end{subfigure}
    \caption{Aim~1 (Continuous outcome): Comparison of CV and computational time across search engines \texttt{gp}, \texttt{bisection} and \texttt{gp-bs} for $R^2$ metric under varying $(\kappa = 10, 20)$, $R^2 = 0.2, 0.7$, number of predictors $(p = 10, 100)$ and total simulation budget $(B = 200$--$2000)$.}
    \label{fig:aim1-continuous-summary2}
\end{figure}
 
\begin{figure}[H]
    \centering
    \begin{subfigure}[t]{0.95\linewidth}
        \centering
        \includegraphics[width=\linewidth]{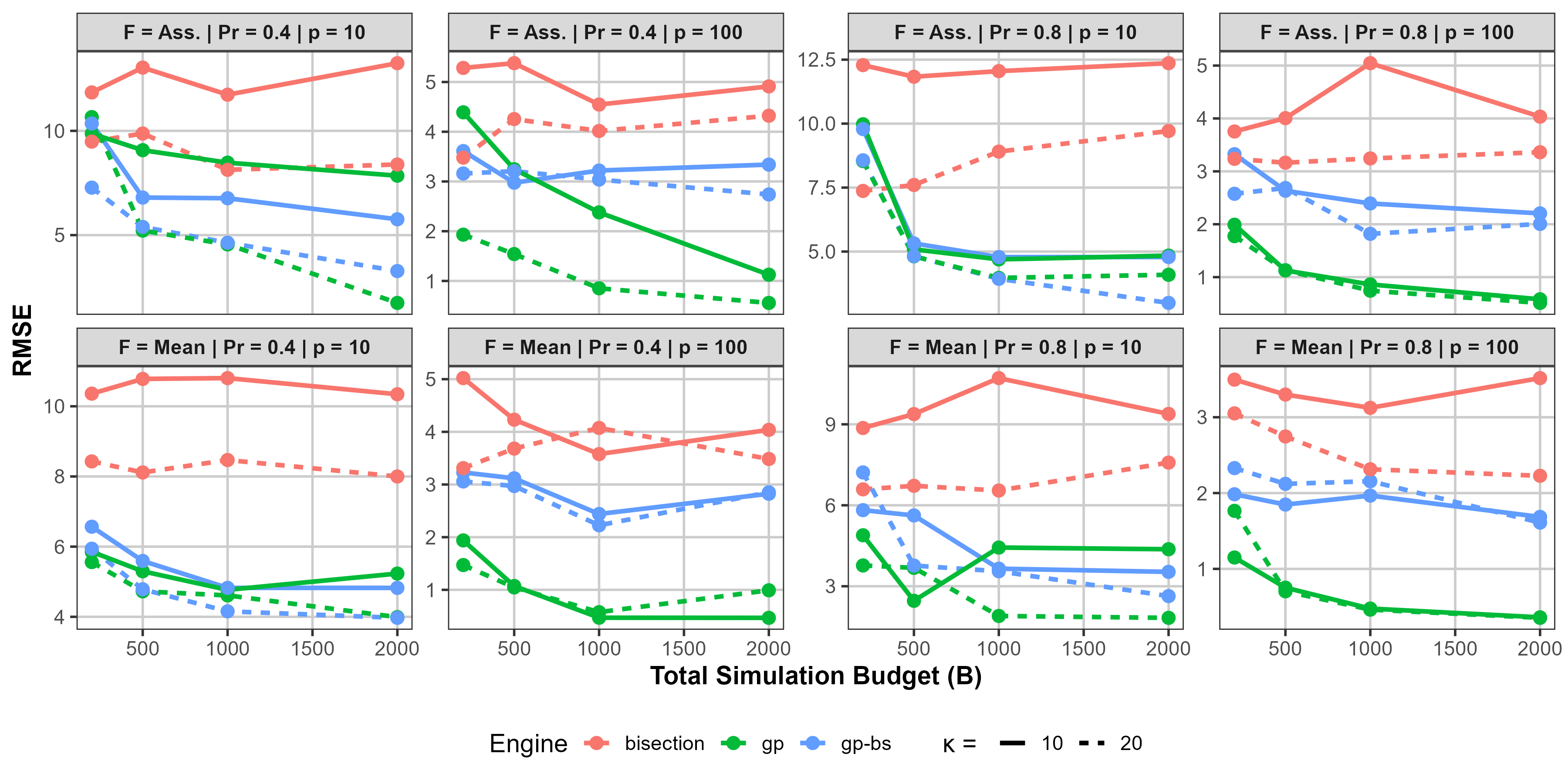}
        \caption{Coefficient of Variation (CV) of sample size estimates relative to mean $n^*$.}
        \label{fig:aim1-surv-cv2}
    \end{subfigure}
    \hspace{0.5cm}
    \begin{subfigure}[t]{0.95\linewidth}
        \centering
        \includegraphics[width=\linewidth]{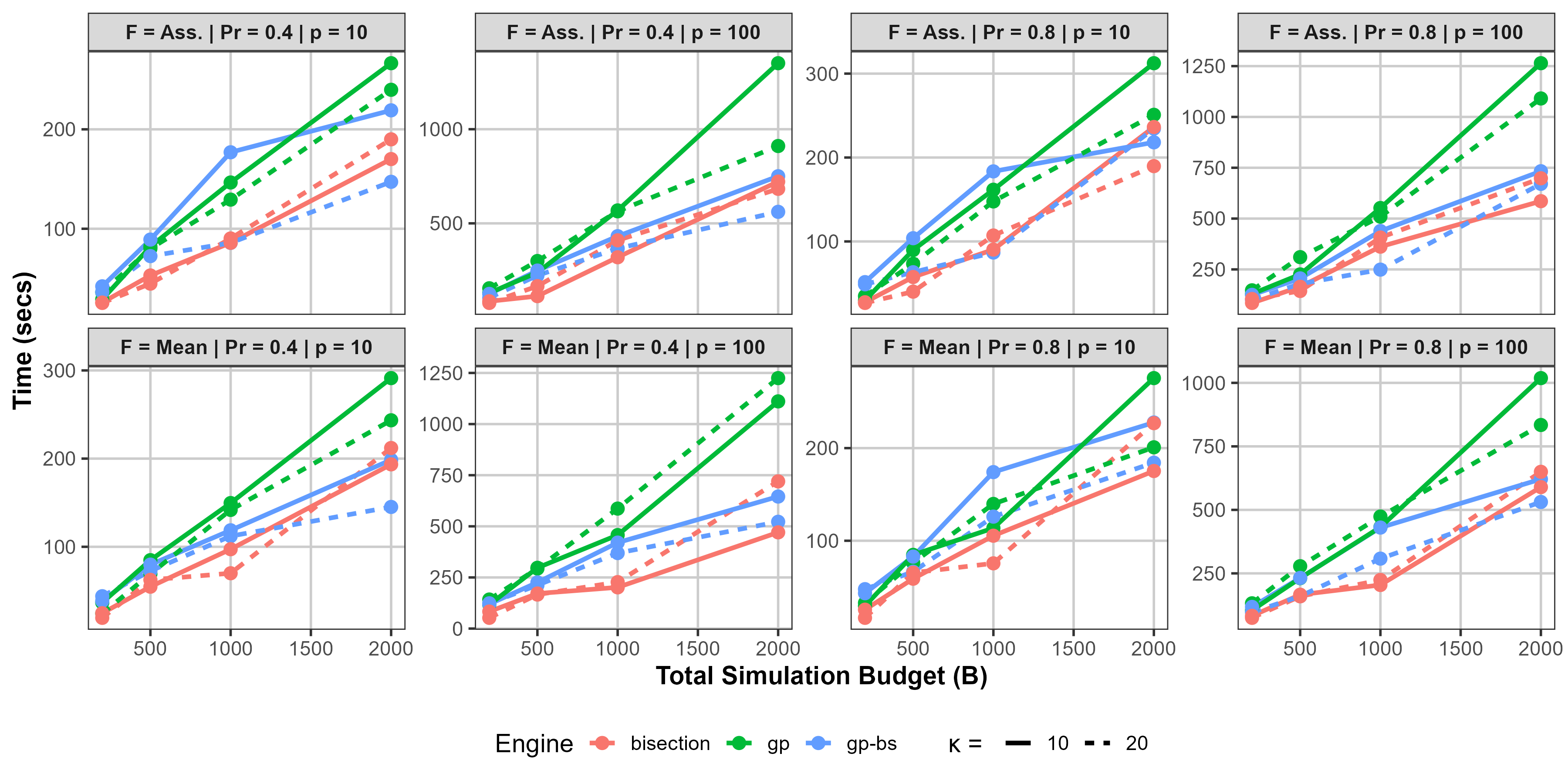}
        \caption{Computational time required to estimate minimum sample size $n^*$.}
        \label{fig:aim1-surv-time2}
    \end{subfigure}
    \caption{Aim~1 (Survival outcome): Comparison of CV and computational time across search engines \texttt{gp}, \texttt{bisection} and \texttt{gp-bs} for C-index metric under varying $(\kappa = 10, 20)$, event rate $(0.4, 0.8)$, number of predictors $(p = 10, 100)$ and total simulation budget $(B = 200$--$2000)$.}
    \label{fig:aim1-survival-summary2}
\end{figure}

\subsection*{Aim~2 Supplementary Figures}
 
\begin{figure}[H]
\centering
\includegraphics[width=\linewidth]{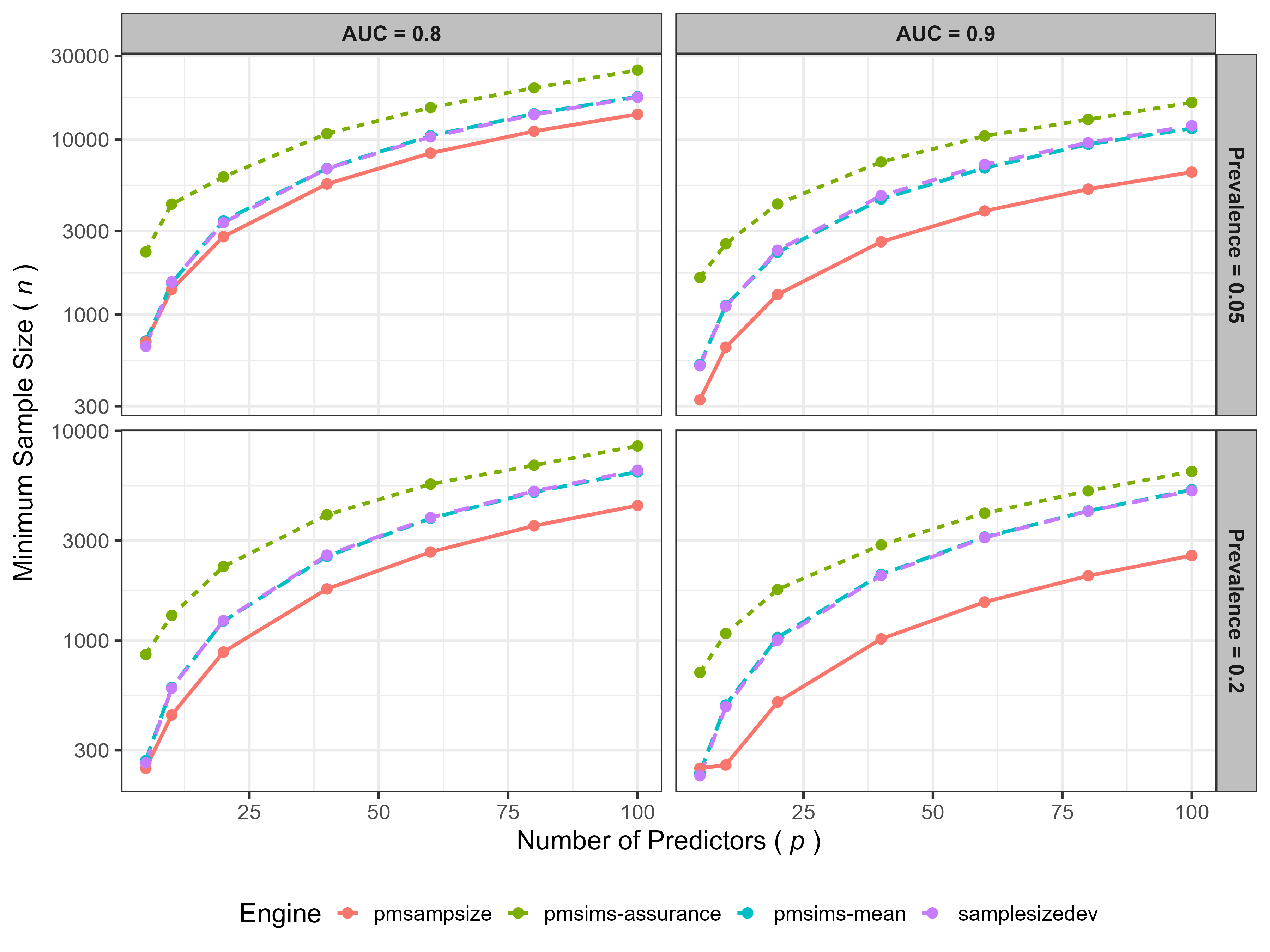}
\caption{Minimum sample size requirements recommended by different engines as a function of the number of predictors, stratified by outcome prevalence and large-sample AUC for binary outcome models targeting a calibration slope of $0.90$.}
\label{fig:aim2-min-n-bin}
\end{figure}
 
\begin{figure}[H]
        \centering
        \includegraphics[width=\linewidth]{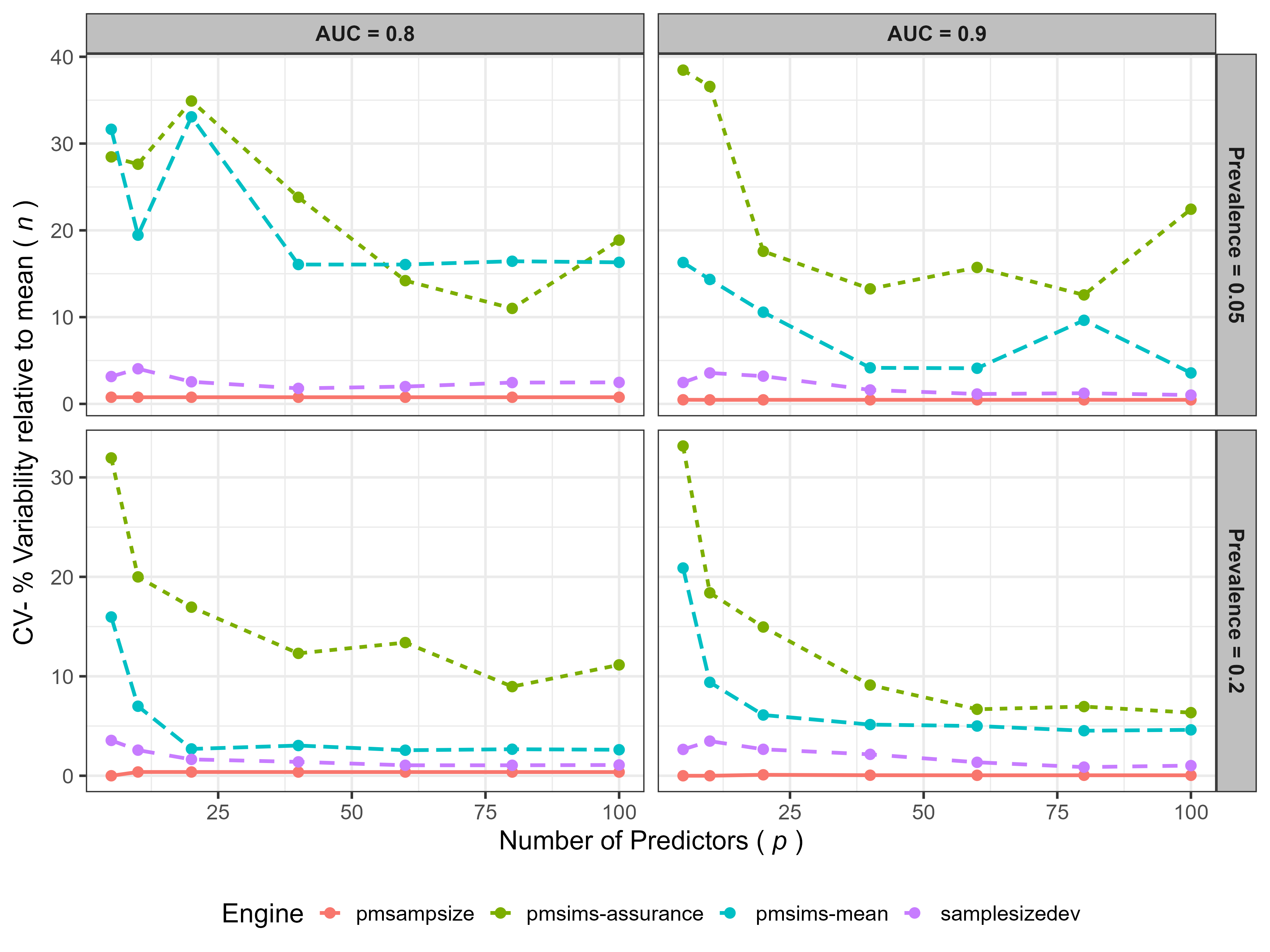}
        \caption{Comparison of relative coefficient of variation in recommended sample sizes across sample size determination engines as a function of the number of candidate predictors, stratified by outcome prevalence and large-sample AUC in binary prediction models.}
        \label{fig:aim2-CV-bin}
\end{figure}
 
\begin{figure}[H]
        \centering
        \includegraphics[width=\linewidth]{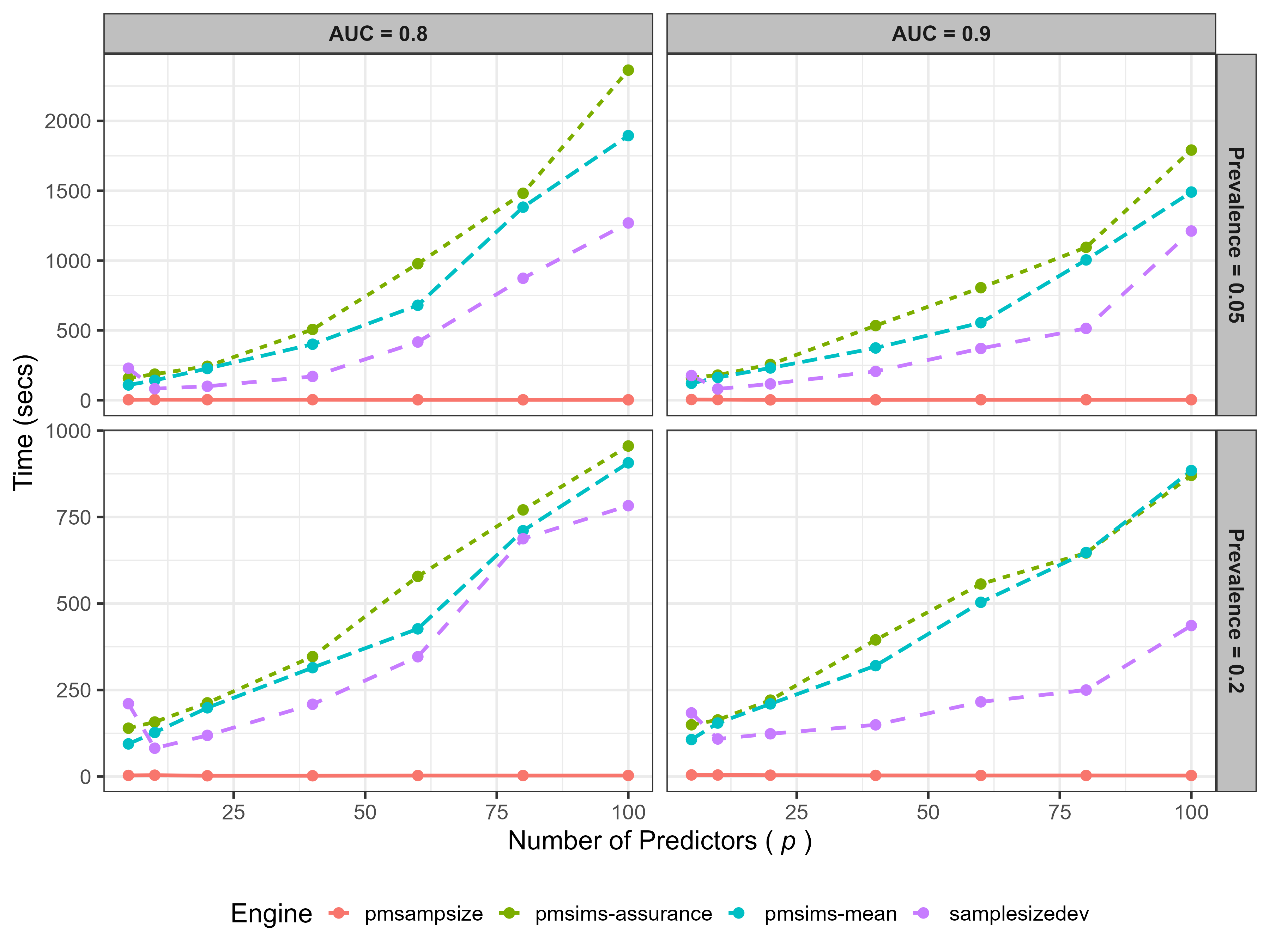}
        \caption{Computational time required by each sample size determination engine as a function of the number of predictors, stratified by outcome prevalence and large-sample AUC in binary prediction models.}
        \label{fig:aim2-time-bin}
\end{figure}
 
\begin{figure}[H]
        \centering
        \includegraphics[width=\linewidth]{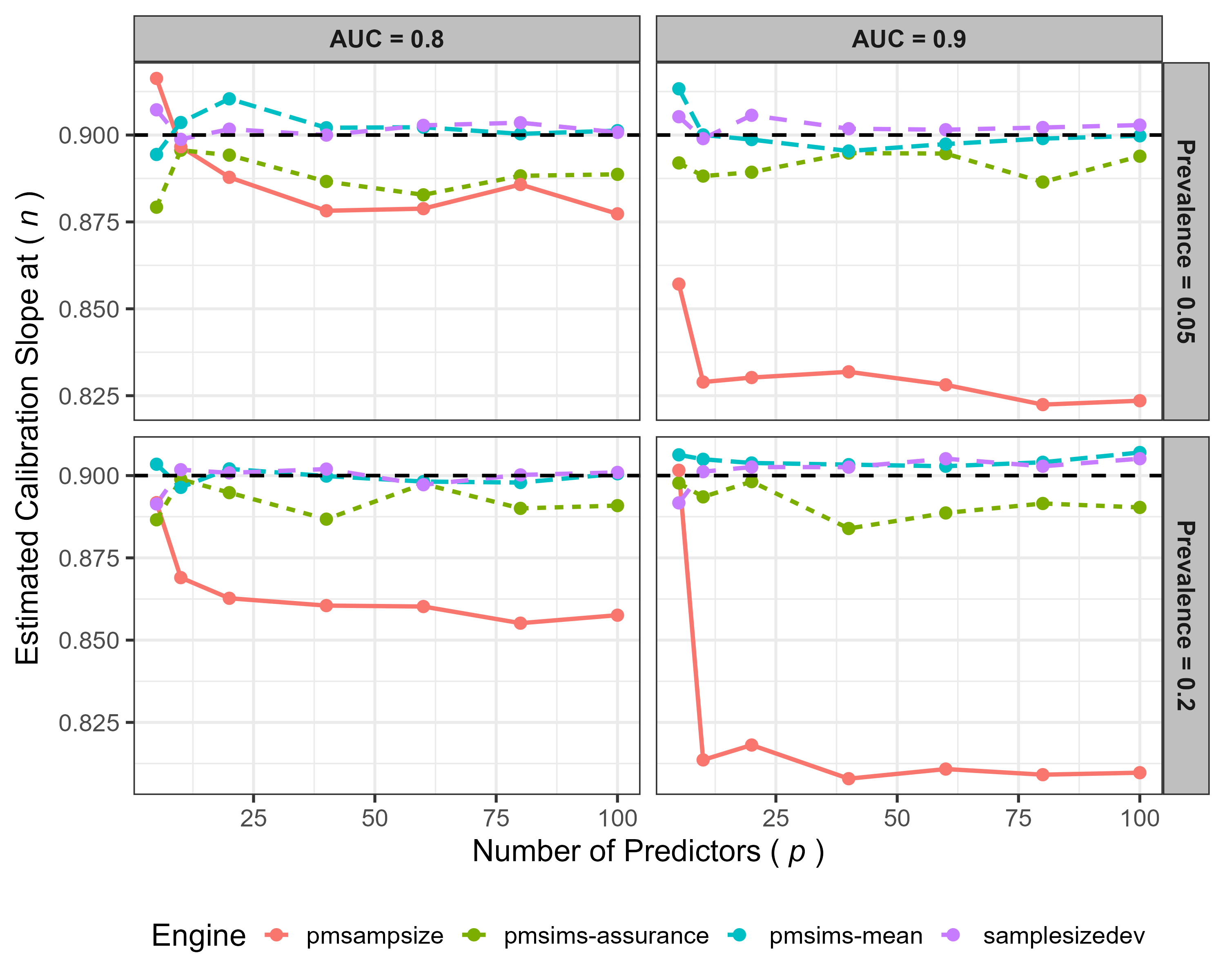}
        \caption{Achieved calibration slope at the recommended sample size for each engine as a function of the number of predictors, stratified by outcome prevalence and large-sample AUC in binary prediction models (dashed line indicates target slope of $0.90$).}
        \label{fig:aim2-perf-bin}
\end{figure}
 
\begin{figure}[H]
        \centering
        \includegraphics[width=\linewidth]{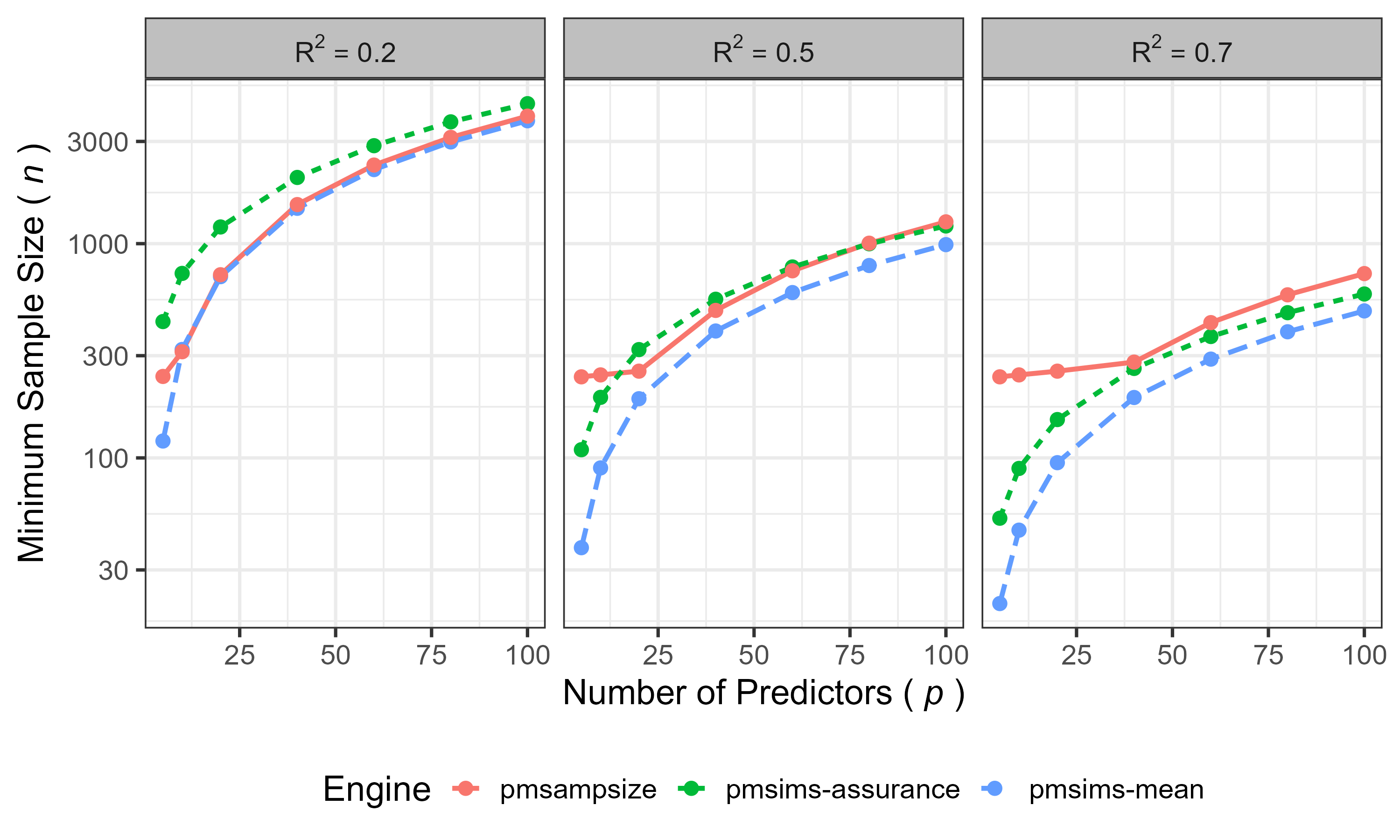}
        \caption{Minimum sample size requirements recommended by different engines as a function of the number of predictors, stratified by large-sample $R^2$ for continuous outcome models targeting a calibration slope of $0.90$.}
        \label{fig:aim2-min-n-cont}
\end{figure}
 
\begin{figure}[H]
        \centering
        \includegraphics[width=.97\linewidth]{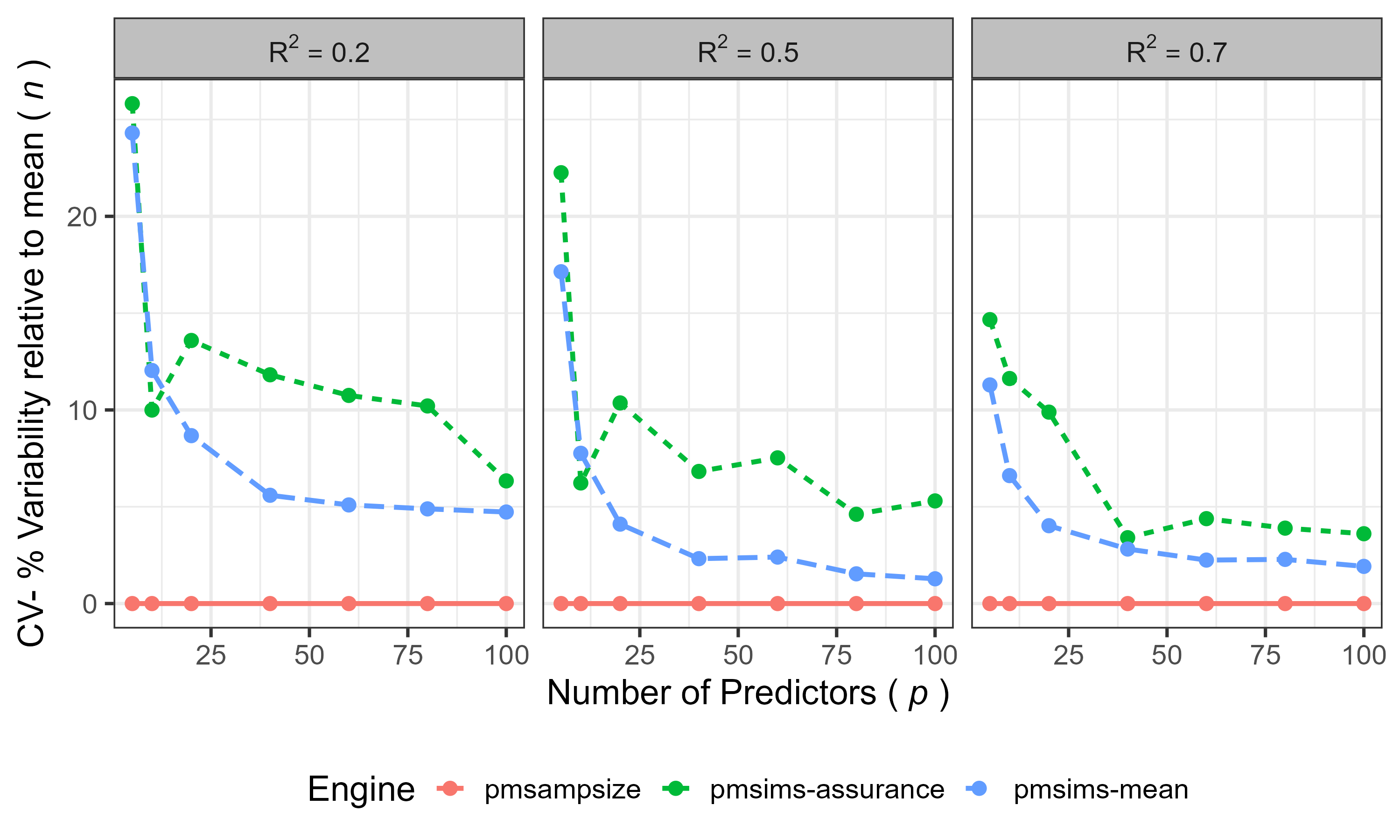}
        \caption{Comparison of relative coefficient of variation in recommended sample sizes across sample size determination engines as a function of the number of candidate predictors, stratified by large-sample $R^2$ in continuous prediction models.}
        \label{fig:aim2-CV-cont}
\end{figure}
 
\begin{figure}[H]
        \centering
        \includegraphics[width=.97\linewidth]{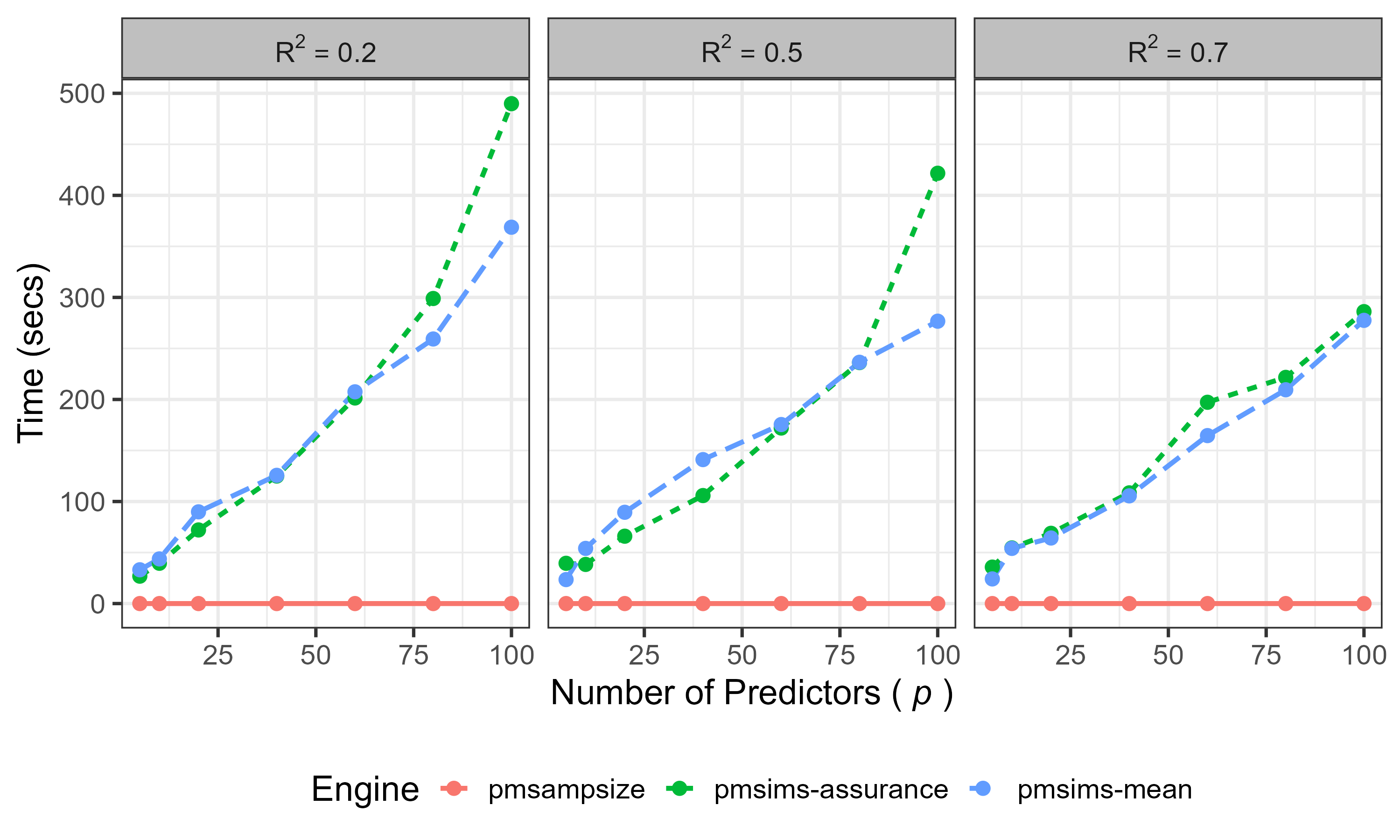}
        \caption{Computational time required by each sample size determination engine as a function of the number of predictors, stratified by large-sample $R^2$ in continuous prediction models.}
        \label{fig:aim2-time-cont}
\end{figure}
 
\begin{figure}[H]
        \centering
        \includegraphics[width=\linewidth]{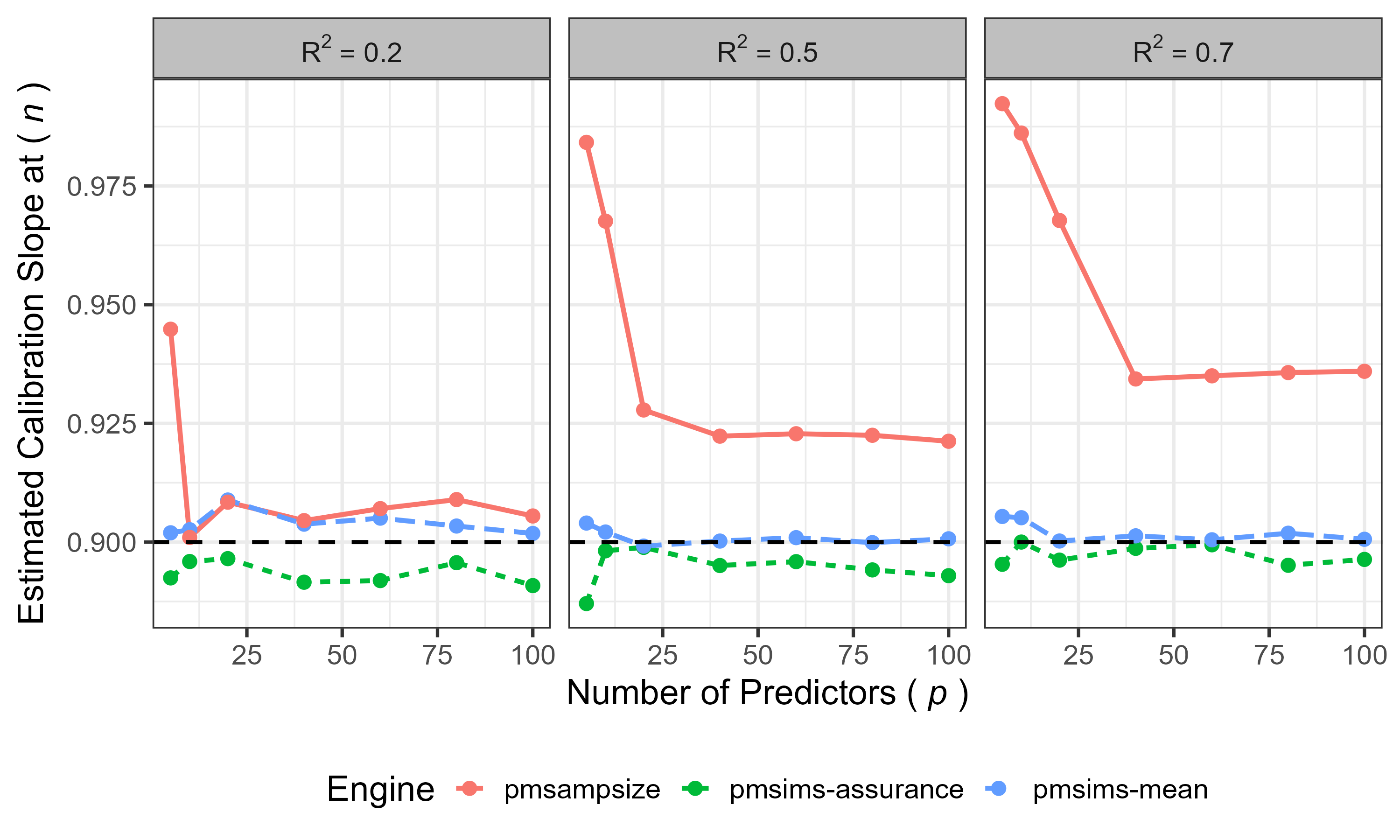}
        \caption{Achieved calibration slope at the recommended sample size for each engine as a function of the number of predictors, stratified by large-sample $R^2$ (dashed line indicates target slope of $0.90$).}
        \label{fig:aim2-perf-cont}
\end{figure}
 
\begin{figure}[H]
        \centering
        \includegraphics[width=\linewidth]{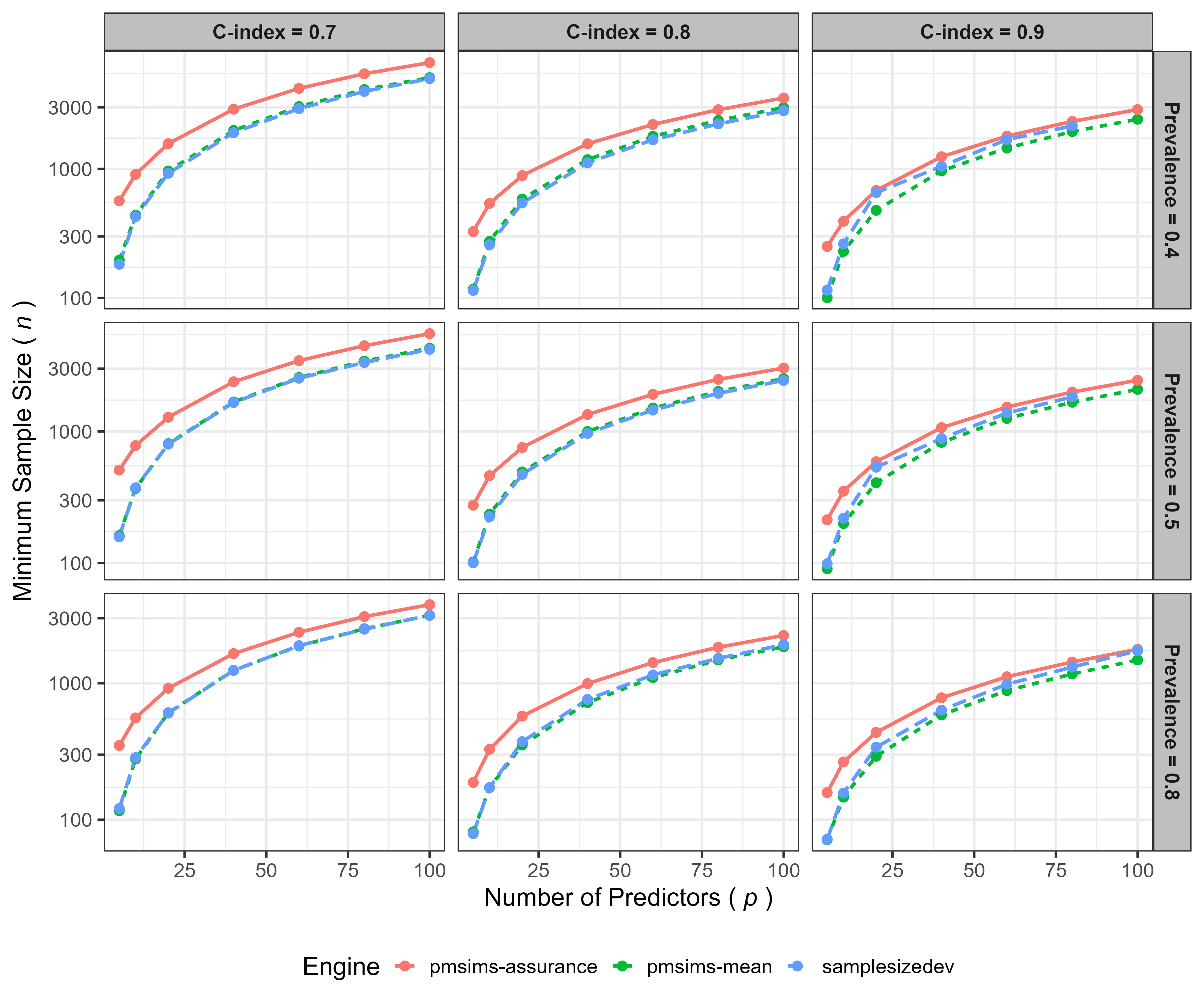}
        \caption{Minimum sample size requirements recommended by different engines as a function of the number of predictors, stratified by event rate and large-sample $C$-index for survival outcome models targeting a calibration slope of $0.90$.}
        \label{fig:aim2-min-n-surv}
\end{figure}
 
\begin{figure}[H]
        \centering
        \includegraphics[width=\linewidth]{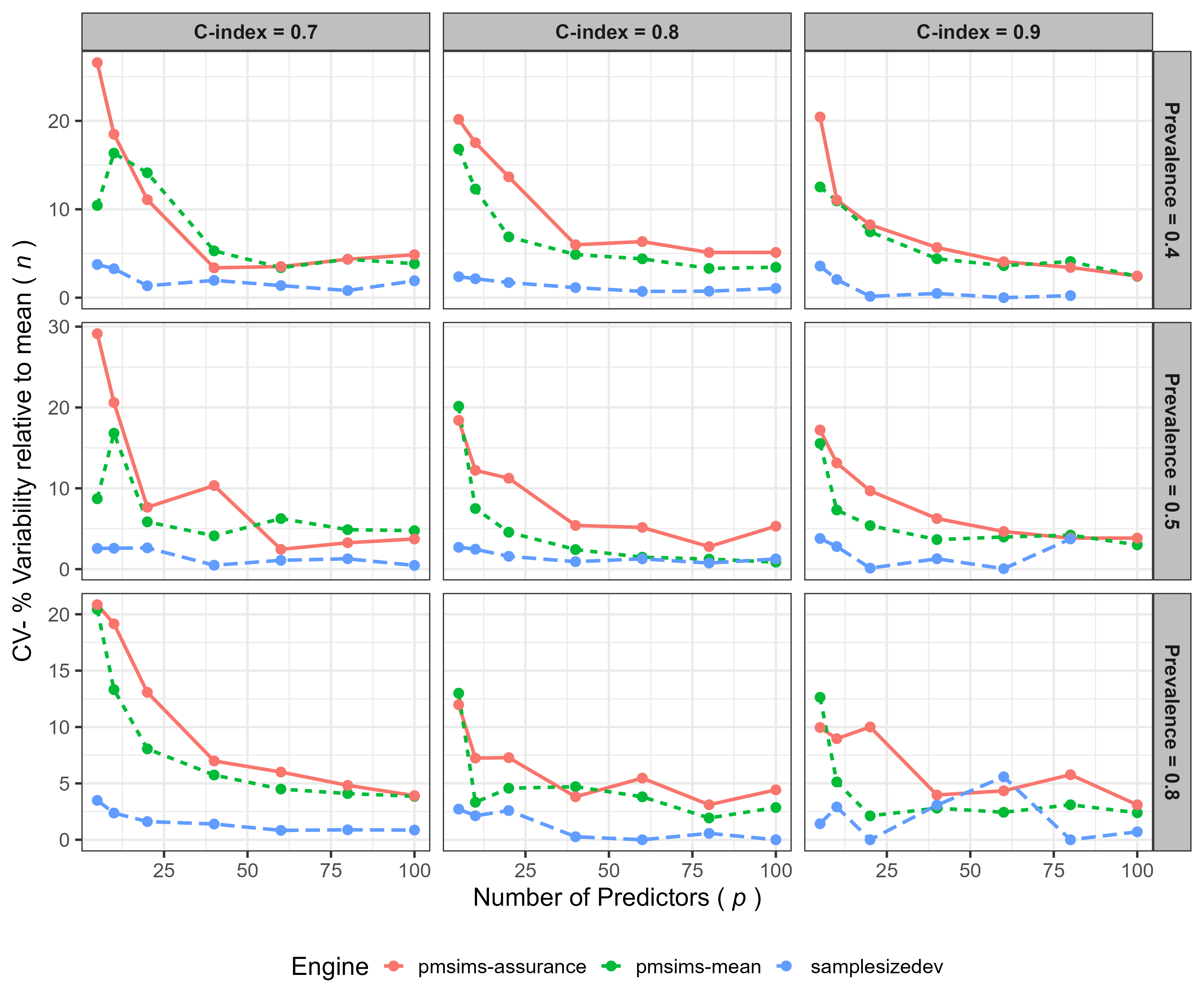}
        \caption{Comparison of relative coefficient of variation in recommended sample sizes across sample size determination engines as a function of the number of candidate predictors, stratified by event rate and large-sample $C$-index in survival prediction models.}
        \label{fig:aim2-CV-surv}
\end{figure}
 
\begin{figure}[H]
        \centering
        \includegraphics[width=\linewidth]{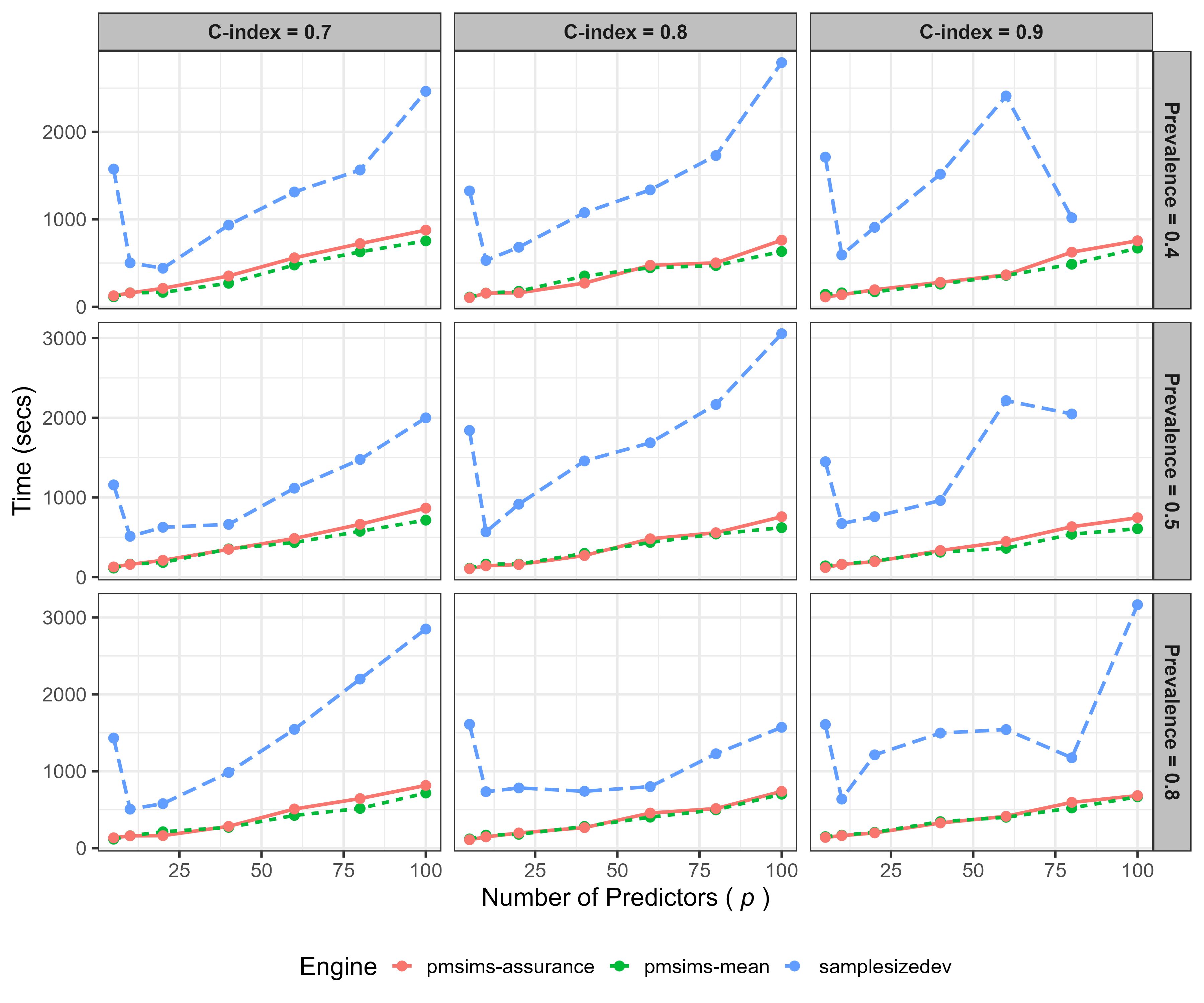}
        \caption{Computational time required by each sample size determination engine as a function of the number of predictors, stratified by event rate and large-sample $C$-index in survival prediction models.}
        \label{fig:aim2-time-surv}
\end{figure}
 
\begin{figure}[H]
        \centering
        \includegraphics[width=\linewidth]{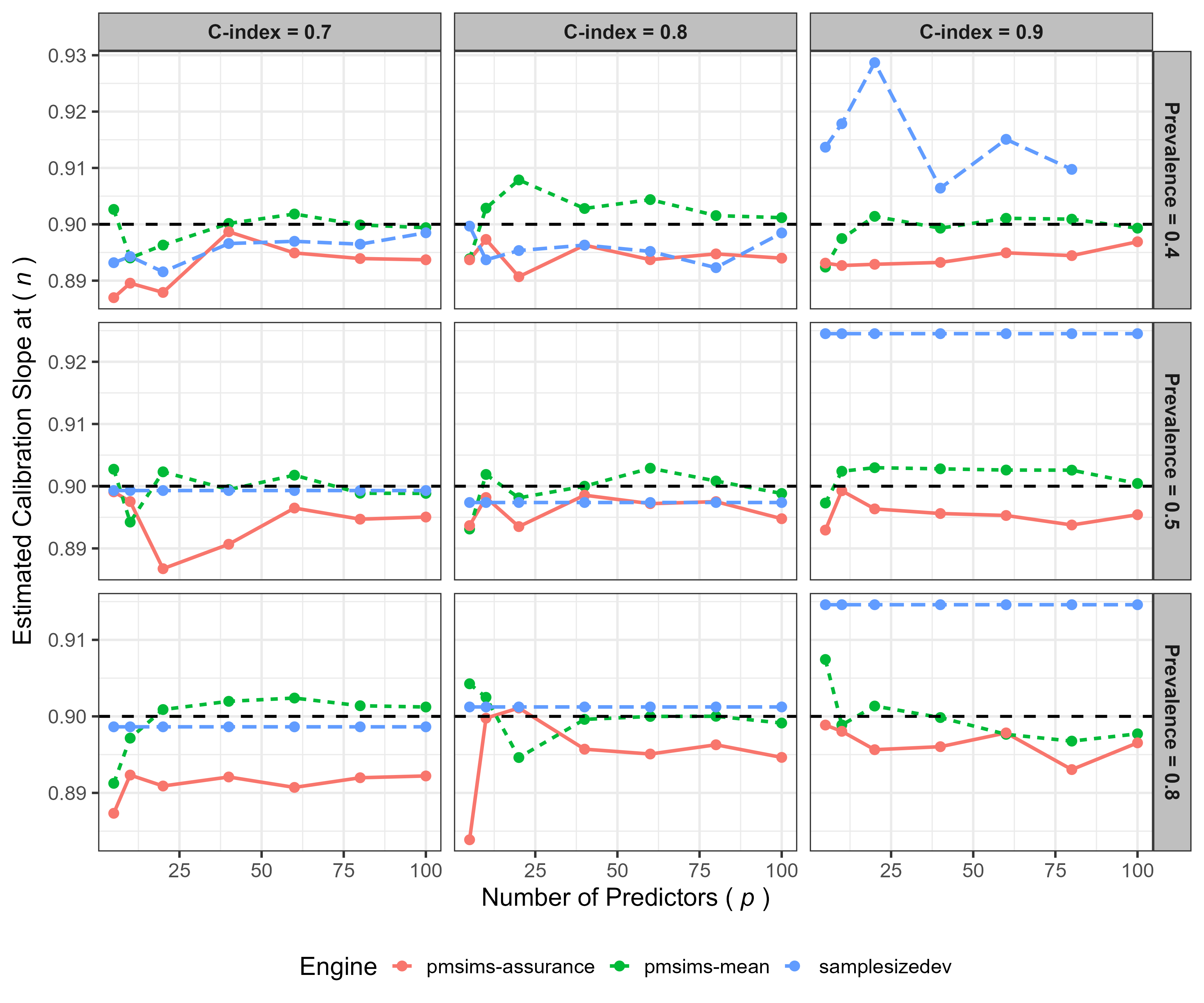}
        \caption{Achieved calibration slope at the recommended sample size for each engine as a function of the number of predictors, stratified by event rate and large-sample $C$-index in survival prediction models (dashed line indicates target slope of $0.90$).}
        \label{fig:aim2-perf-surv}
\end{figure}
 
\end{appendices}

\bibliography{sn-bibliography}

\end{document}